\documentclass[aps,prc,showpacs,showkeys,nofootinbib]{revtex4}

\usepackage{graphicx}
\usepackage{dcolumn}
\usepackage{amsthm}
\usepackage{bm}

\newcommand{\definition}{\textit}

\newcommand{\sigmabf}{\mbox{\boldmath $\sigma$}}
\newcommand{\xibf}{\mbox{\boldmath $\xi$}}
\newcommand{\rhobf}{\mbox{\boldmath $\rho$}}
\newcommand{\rhobfsm}{\small \mbox{\boldmath $\rho$}}

\begin{document}

\title{New approach to determine proton-nucleus interactions from experimental bremsstrahlung data}

\author{Sergei~P.~Maydanyuk}%
\email{maidan@kinr.kiev.ua}%
\affiliation{Institute of Modern Physics, Chinese Academy of Sciences,
Lanzhou, 730000, China}
\affiliation{Institute for Nuclear Research, National Academy of Sciences
of Ukraine, Kiev, 03680, Ukraine}
\author{Peng-Ming~Zhang}%
\email{zhpm@impcas.ac.cn} %
\affiliation{Institute of Modern Physics, Chinese Academy of Sciences,
Lanzhou, 730000, China}

\date{\small\today}
\begin{abstract}
A new approach is presented to determine the proton-nucleus interactions from the analysis of the accompanying photon bremsstrahlung.
We study the scattering of $p + ^{208}{\rm Pb}$ at the proton incident energies of 140 and 145~MeV,
and the scattering of $p + ^{12}{\rm C}$, $p + ^{58}{\rm Ni}$, $p + ^{107}{\rm Ag}$ and $p + ^{197}{\rm Au}$ at the proton incident energy of 190~MeV.
The model determines contributions of the coherent emission (formed by an interaction between the scattering proton and nucleus as a whole without the internal many-nucleon structure),
incoherent emission (formed by interactions between the scattering proton and
nucleus with the internal many-nucleon structure),
and transition between them in dependence on the photon energy.
The radius-parameter of the proton-nucleus potential for these reactions is extracted from the experimental bremsstrahlung data analysis.
We explain the hump-shaped plateau in the intermediate- and high-energy regions of the spectra by the essential presence of the incoherent emission,
while at low energies the coherent emission predominates which produces the logarithmic shape spectrum.
We provide our predictions (in absolute scale) for the angular distribution of the bremsstrahlung photons in order to test our model, results and analysis in further experiments.
\end{abstract}



\maketitle

\section{Introduction
\label{sec.inroduction}}

The optical model has a significant impact on many branches of nuclear reaction physics. In frameworks of such a model, our understanding about interactions between two colliding nuclear fragments is based on the agreement between experimental and calculated cross sections.
In particular, all possible physical aspects are incorporated into the model in order to fit the experimental data, including the different forms of interactions between nucleons, many-nucleons aspects, dynamic approaches, nonlocal quantum properties, etc.
The applied numerical techniques, chosen approximations, and imposed boundary conditions are important for the resulting calculations of the cross sections.
However, different input parameters, which correspond to quite different physical pictures, may lead to similar final results (i.e. cross-sections). This indicates uncertainties to determine the parameters of the potentials of the nucleon-nucleus and nucleus-nucleus interactions.

Because of this, it is interesting to find an alternate way to extract the information about the interacting potentials. Here, the bremsstrahlung emission of photons accompanying the scattering of protons off nuclei attracts a lot of attention. The cross sections of the emitted photons have been measured for a long time
(see Refs.~\cite{Edington.1966.NP,Koehler.1967.PRL,Kwato_Njock.1988.PLB,Pinston.1989.PLB,
Pinston.1990.PLB,Clayton.1992.PRC,Pluiko.1987.PEPAN,Kamanin.1989.PEPAN,Clayton.1991.PhD,
Chakrabarty.1999.PRC,Goethem.2002.PRL}),
and different theoretical models and approaches were developed to estimate the emitted photons (for example,
see \cite{Nakayama.1986.PRC,Nakayama.1989.PRC,Nakayama.1989.PRCv40,Knoll.1989.NPA,Herrmann.1991.PRC,Liou.1987.PRC,Liou.1993.PRC,Liou.1995.PLB.v345,%
Liou.1996.PRC,Li.1998.PRC.v57,Li.1998.PRC.v58,Timmermans.2001.PRC,Liou.2004.PRC,%
Li.2005.PRC,Timmermans.2006.PRC,Li.2011.PRC,Maydanyuk.2011.JPG,Kurgalin.2001.IRAN}).
In particular, the spectra of the emitted photons are dependent on dynamics of the scattering of the proton off the nucleus, which is determined by interactions between the proton and nucleus. However, until now it has been unclear how information about such interactions could be extracted from the bremsstrahlung spectra analysis.

The problem is that high accuracy in calculations is required to determine parameters of the interacting proton-nucleus potentials; however, the convergence of such calculations is extremely low.
Such a problem was noted previously in \cite{Kopitin.1997.YF} where the authors of that paper performed calculations with realistic interactions between the nucleon and nucleus. The additional indication is absent of any clear information in the literature about the determination of parameters of the potential by this approach, while the history of the bremsstrahlung research is extremely long.
In this paper we develop such an approach to the problem of the scattering of proton off nucleus.

We start our analysis from the $p + ^{208}{\rm Pb}$ reaction, which was intensively studied by different research groups and,
thus has more evidence~\cite{Edington.1966.NP,Clayton.1991.PhD,Clayton.1992.PRC}.
The authors of~\cite{Clayton.1991.PhD,Clayton.1992.PRC} clearly observed the hump-shaped plateau in the experimental spectrum which is different from the typical exponential shape of the bremsstrahlung spectra previously measured by Edington and Rose in \cite{Edington.1966.NP}.
The further careful measurements of the bremsstrahlung emission in the proton nucleus scattering were done by the TAPS collaboration \cite{Goethem.2002.PRL} and results confirmed the clear presence of such a plateau in the spectra.
A supposition to explain such behavior of the spectra is to consider the internal dynamic motion of nucleons and collisions between them.
Here, Nakayama and Bertsch indicate an important role of the individual nucleon-nucleon interactions in the proton-nucleus bremsstrahlung (see Refs.~\cite{Nakayama.1986.PRC,Nakayama.1989.PRC,Nakayama.1989.PRCv40}).
However, measurements of emission of the bremsstrahlung photons in the $\alpha$ decay show the
absence of such a hump-shaped plateau in the spectra in the $\alpha$ decay (see \cite{Boie.2007.PRL,Boie.2009.PhD} for details).
Co-existence of two different types of emission of photons requires more careful consideration of the internuclear processes inside the nucleus which could form an emission of photons.
In order to clarify these questions, many-nucleon structure of the nucleus is included in our model and analysis.

\section{Model
\label{sec.2}}

\subsection{Generalized Pauli equation for many-nucleon system
\label{sec.2.1}}

We shall start from a generalization of the Pauli equation
on the system composed from $A+1$ nucleons,
describing scattering of proton off nucleus with $A$ nucleons,
where the Hamiltonian can be constructed as \cite{Maydanyuk.2012.PRC}
\begin{equation}
\begin{array}{lcl}
  \hat{H} =
  \displaystyle\sum_{i=1}^{A+1}
  \biggl\{
    \displaystyle\frac{1}{2\,m_{i}}\;
    \Bigl( \mathbf{p}_{i} - \displaystyle\frac{z_{i}e}{c} \mathbf{A}_{i} \Bigr)^{2} -
    \displaystyle\frac{z_{i}e\hbar}{2m_{i}c}\, \sigmabf \cdot \mathbf{rot A}_{i} +
    z_{i}e\, A_{i,0}
  \biggr\} +
  V(\mathbf{r}_{1} \ldots \mathbf{r}_{A+1}) = \hat{H}_{0} + \hat{H}_{\gamma},
\end{array}
\label{eq.2.1.1}
\end{equation}
where
\begin{equation}
\begin{array}{lcl}
  \hat{H}_{0} =
  \displaystyle\sum_{i=1}^{A+1}
    \displaystyle\frac{1}{2\,m_{i}}\: \mathbf{p}_{i}^{2} +
  V(\mathbf{r}_{1} \ldots \mathbf{r}_{A+1}), \\
  \hat{H}_{\gamma} =
  \displaystyle\sum_{i=1}^{A+1}
  \biggl\{
    - \displaystyle\frac{z_{i} e}{m_{i}c}\; \mathbf{p}_{i} \cdot \mathbf{A}_{i} +
    \displaystyle\frac{z_{i}^{2}e^{2}}{2m_{i}c^{2}} \mathbf{A}_{i}^{2} -
    \displaystyle\frac{z_{i}e\hbar}{2m_{i}c}\, \sigmabf \cdot \mathbf{rot A}_{i} +
    z_{i}e\, A_{i,0}
  \biggr\}.
\end{array}
\label{eq.2.1.2}
\end{equation}
Here, $m_{i}$ and $z_{i}$ are the mass and electromagnetic charge of
nucleon with number $i$,
$\mathbf{p}_{i} = -i\hbar\, \mathbf{d}/\mathbf{dr}_{i} $ is the momentum
operator for a nucleon with number $i$,
$V(\mathbf{r}_{1} \ldots \mathbf{r}_{A+1})$ is the general form of the
potential of interactions between nucleons,
$\sigmabf$ are Pauli matrices,
$A_{i} = (\mathbf{A}_{i}, A_{i,0})$ is the potential of the electromagnetic
field formed by moving a nucleon with number $i$.
Let us turn to the center-of-mass frame.
Introducing a coordinate of centers of masses for the nucleus
$\mathbf{R}_{A} = \sum_{j=1}^{A} m_{j}\, \mathbf{r}_{A j} / m_{A}$,
coordinate of centers of masses of the complete system
$\mathbf{R} = (m_{A}\mathbf{R}_{A} + m_{\rm p}\mathbf{r}_{\rm p}) / (m_{A}+m_{\rm p})$,
relative coordinates $\rhobf_{A j} = \mathbf{r}_{j} - \mathbf{R}_{A}$ and
$\mathbf{r} = \mathbf{r}_{\rm p} - \mathbf{R}_{A}$,
we obtain new independent variables $\mathbf{R}$, $\mathbf{r}$ and
$\rhobf_{Aj}$ ($j=1 \top\cdots A-1$)
\begin{equation}
\begin{array}{lll}
  \mathbf{R} =
    \displaystyle\frac{1}{m_{A}+m_{\rm p}}\;
    \Bigl\{
      \displaystyle\sum_{j=1}^{A} m_{Aj}\, \mathbf{r}_{A j} +
      m_{\rm p}\, \mathbf{r}_{\rm p}
    \Bigr\}, &
  \mathbf{r} =
  \mathbf{r}_{\rm p} - \displaystyle\frac{1}{m_{A}} \displaystyle\sum_{j=1}^{A}
  m_{Aj}\, \mathbf{r}_{Aj}, &
  \rhobf_{A j} =
  \mathbf{r}_{A j} -
  \displaystyle\frac{1}{m_{A}} \displaystyle\sum_{k=1}^{A} m_{Ak}\, \mathbf{r}_{Ak},
\end{array}
\label{eq.2.1.3}
\end{equation}
and calculate operators of corresponding momenta
\begin{equation}
\begin{array}{lclll}
  \mathbf{p}_{\rm p} =
    - i\hbar\, \displaystyle\frac{\mathbf{d}}{\mathbf{dr}_{\rm p}} =
    \displaystyle\frac{m_{\rm p}}{m_{A} + m_{\rm p}}\, \mathbf{P} + \mathbf{p}, &
  \mathbf{p}_{Aj} =
    - i\hbar\, \displaystyle\frac{\mathbf{d}}{\mathbf{dr}_{Aj}} =
    \displaystyle\frac{m_{Aj}}{m_{A} + m_{\rm p}}\, \mathbf{P} -
    \displaystyle\frac{m_{Aj}}{m_{A}}\,\mathbf{p} +
    \displaystyle\frac{m_{A} - m_{Aj}}{m_{A}}\; \mathbf{\tilde{p}}_{Aj} -
    \displaystyle\frac{m_{Aj}}{m_{A}}\,
    \displaystyle\sum_{k=1, k \ne j}^{A-1} \mathbf{\tilde{p}}_{A k},
\end{array}
\label{eq.2.1.4}
\end{equation}
where
$\mathbf{P} = -i\hbar\, \mathbf{d}/\mathbf{dR}$,
$\mathbf{p} = -i\hbar\, \mathbf{d}/\mathbf{dr}$,
$\mathbf{\tilde{p}}_{Aj} = -i\,\hbar\, \mathbf{d}/\mathbf{d}\rhobf_{Aj}$,
$m_{\rm p}$ and $m_{A}$ are masses of the scattering proton and nucleus.
We find the kinetic term of the unperturbed Hamiltonian
(at an approximation of $\displaystyle\sum_{j=1}^{A} m_{Aj} = m_{A}$):
\begin{equation}
\begin{array}{ll}
  \displaystyle\sum_{i=1}^{A+1}
    \displaystyle\frac{1}{2\,m_{i}}\: \mathbf{p}_{i}^{2} =
  \displaystyle\frac{1}{2\,(m_{A} + m_{\rm p})}\: \mathbf{P}^{2} +
  \displaystyle\frac{m_{A} + m_{\rm p}}{2\,m_{\rm p}\, m_{A}}\: \mathbf{p}^{2} +
  \hat{T}_{\rm nucl}. &
\end{array}
\label{eq.2.1.7}
\end{equation}
The first term on the right-hand side (r. h. s.)
represents the kinetic energy of motion of the full proton-nucleus system,
the second term --- kinetic energy of relative motion of the proton concerning nucleus,
and the last term $\hat{T}_{\rm nucl}$ --- kinetic energy of the
internal motion of nucleons inside nucleus having the form:
\begin{equation}
\begin{array}{ll}
  \hat{T}_{\rm nucl} =
  \displaystyle\sum_{j=1}^{A-1}
    \displaystyle\frac{1}{2\,m_{Aj}}\: \mathbf{\tilde{p}}_{Aj}^{2} -
  \displaystyle\frac{1}{2\,m_{A}}\:
    \biggl\{ \displaystyle\sum_{k=1}^{A-1} \mathbf{\tilde{p}}_{Ak} \biggr\}^{2}.
\end{array}
\label{eq.2.1.8}
\end{equation}


Let us study the leading emission operator of the system composed of the proton
and nucleus in the laboratory frame:
\begin{equation}
\begin{array}{lcl}
  \hat{H}_{\gamma} =
    - \displaystyle\frac{z_{\rm p}\,e}{m_{\rm p}c}\;
    \mathbf{A}(\mathbf{r}_{\rm p},t) \cdot \mathbf{\hat{p}}_{\rm p} -
    \displaystyle\sum\limits_{j=1}^{A}
    \displaystyle\frac{z_{j}\,e}{m_{j}c}\;
    \mathbf{A}(\mathbf{r}_{j},t) \cdot \mathbf{\hat{p}}_{j}.
\end{array}
\label{eq.2.2.1}
\end{equation}
Here, $\mathbf{A}(\mathbf{r}_{s},t)$ describes emission of photon
caused by nucleon with number $s$ ($s=p$ is for proton, $s=j$ for
nucleons of nucleus).
Using its presentation in form~(5) of~\cite{Maydanyuk.2012.PRC},
for the emission operator 
in the center-of-mass frame we obtain
\begin{equation}
\begin{array}{lcl}
  \hat{H}_{\gamma} =
  -\,e\, \sqrt{\displaystyle\frac{2\pi\hbar}{w_{\rm ph}}}\,
    \displaystyle\sum\limits_{\alpha=1,2} \mathbf{e}^{(\alpha),*}\;
    e^{-i \mathbf{k} \cdot \bigl[\mathbf{R} - \displaystyle\frac{m_{\rm p}}{M+m_{\rm p}}\:
    \mathbf{r}\bigr]} \cdot
  \Biggl\{
    \displaystyle\frac{1}{M + m_{\rm p}}\:
    \biggl[
      e^{-i \mathbf{k} \cdot \mathbf{r}}\; z_{\rm p} +
      \displaystyle\sum\limits_{j=1}^{A} z_{Aj}\; e^{-i \mathbf{k} \cdot \rhobfsm_{Aj}}
    \biggr]\; \mathbf{P}\; +\; \\
  + \;
    \biggl[
      e^{-i \mathbf{k} \cdot \mathbf{r}}\; \displaystyle\frac{z_{\rm p}}{m_{\rm p}}\; -
      \displaystyle\frac{1}{M}
      \displaystyle\sum\limits_{j=1}^{A}
        z_{Aj}\; e^{-i \mathbf{k} \cdot \rhobfsm_{Aj}}
    \biggr]\; \mathbf{p}\; + \;
  \displaystyle\sum\limits_{j=1}^{A-1}
      \displaystyle\frac{z_{Aj}}{m_{Aj}}\; e^{-i \mathbf{k} \cdot
      \rhobfsm_{Aj}}\, \mathbf{\tilde{p}}_{Aj}\; -\;
  \displaystyle\frac{1}{M}
  \biggl[
    \displaystyle\sum\limits_{j=1}^{A}
      z_{Aj}\; e^{-i \mathbf{k} \cdot \rhobfsm_{Aj}}
  \biggr]\; \displaystyle\sum_{k=1}^{A-1} \mathbf{\tilde{p}}_{Ak}
  \Biggr\},
\end{array}
\label{eq.2.2.2}
\end{equation}
where the star denotes the complex conjugation,
$\mathbf{e}^{(\alpha)}$ are unit vectors of the polarization of the photon emitted
($\mathbf{e}^{(\alpha),*} = \mathbf{e}^{(\alpha)}$), $\mathbf{k}$ is the wave vector of the photon, and
$w_{\rm ph} = k\, c = \bigl| \mathbf{k}\bigr|\, c$.
Vectors $\mathbf{e}^{(\alpha)}$ are perpendicular to $\mathbf{k}$ in the Coulomb gauge.
We have two polarizations $\mathbf{e}^{(1)}$ and $\mathbf{e}^{(2)}$
for the photon with momentum $\mathbf{k}$ ($\alpha = 1,2$)
with properties~\cite{Maydanyuk.2012.PRC}
\begin{equation}
\begin{array}{lclc}
  \Bigl[ \mathbf{k}_{\rm ph} \times \mathbf{e}^{(1)} \Bigr] = k_{\rm ph}\, \mathbf{e}^{(2)}, &
  \Bigl[ \mathbf{k}_{\rm ph} \times \mathbf{e}^{(2)} \Bigr] = -\, k_{\rm ph}\, \mathbf{e}^{(1)}, &
  \displaystyle\sum\limits_{\alpha=1,2}
  \Bigl[ \mathbf{k}_{\rm ph} \times \mathbf{e}^{(\alpha)} \Bigr] =
  k_{\rm ph}\, (\mathbf{e}^{(2)} - \mathbf{e}^{(1)}).
\end{array}
\label{eq.2.2.3}
\end{equation}

\subsection{Wave function of the many-nucleon system
\label{sec.2.3}}

The bremsstrahlung of the emitted photons in nuclear reactions was previously studied, with nuclei described at the microscopic
level~\cite{Baye.1985.NPA,Baye.1991.NPA,Liu.1990.PRC,Dohet-Eraly.2013.PhD,Dohet-Eraly.2011.PRC,Dohet-Eraly.2011.JPCS,Dohet-Eraly.2013.JPCS,Dohet-Eraly.2013.PRC}.
%
%
However, such a formalism was mainly oriented on the nuclear systems with smaller numbers of nucleons.
At the same time, we wish to include heavy nuclei in the analysis, and to use our previous quantum developments (where our description of existing experimental data for proton-nucleus scattering was the most accurate in comparison with other approaches, see~\cite{Maydanyuk.2012.PRC} for details) so we have developed a new approach.

Emission of the bremsstrahlung photons is caused by the relative motion of nucleons of the full nuclear system. However, we assume that the most intensive emission of photons is formed by the relative motion of a proton concerning the nucleus. So, it is sensible to represent the total wave function via coordinates of relative motion of these complicated objects.
Following such logic, we define the wave function of the full nuclear system as
\begin{equation}
\begin{array}{lcl}
  \Psi_{s} =
  \Phi_{s} (\mathbf{R}) \cdot
  \Phi_{\rm p-nucl, s}(\mathbf{r}) \cdot
  \psi_{\rm nucl, s}(\beta_{A}), &
  \Phi_{s} (\mathbf{R}) =  N_{s}\; e^{-i\,\mathbf{K}_{s}\cdot\mathbf{R}}, 
\end{array}
\label{eq.2.3.1}
\end{equation}
where
$s = i$ or $f$ (indexes $i$ and $f$ denote the initial state, i.e. the state before emission of the photon,
and the final state, i.e. the state after emission of the photon),
$\mathbf{K}_{s}$ is full momentum of the proton--nucleus system (in the laboratory frame),
$\Phi_{s} (\mathbf{R})$ is the wave function describing the motion of the center-of-mass of the full nuclear system in the laboratory frame,
$\Phi_{\rm p-nucl,s} (\mathbf{r})$ is a function describing the relative motion (with tunneling for under-barrier energies) of the proton concerning to nucleus (without a description of internal relative motions of nucleons inside the nucleus),
$\psi_{\rm nucl,s}(\beta)$ is the many-nucleon function describing internal states of nucleons in the nucleus
(it determines space states on the basis of relative distances $\rhobf_{1}$ \ldots $\rhobf_{A}$ of nucleons of the nucleus concerning its center-of-mass, and spin-isospin states also),
$\beta_{A}$ is a set of numbers $1 \ldots A$ of nucleons of the nucleus.
$N_{s}$ is a normalized factor which will be defined later.

The motion of nucleons of the nucleus relative to each other does not influence the states describing the relative motion of a proton concerning the nucleus and, therefore, such a representation of the full wave function can be considered as an approximation.
However, the relative internal motions of nucleons of the nucleus provide their own contributions to the full bremsstrahlung spectrum and they can be estimated.
We shall include many-nucleon structure into the wave function $\psi_{\rm nucl, s}(\beta_{A})$ of the nucleus while we assume that the wave function of relative motion $\psi_{\rm p-nucl,s}(\mathbf{r})$ is calculated without it but with maximal orientation of the proton-nucleus potential well extracted from experimental data of proton-decay and proton-nucleus scattering (many-nucleon corrections in it can be taken into account in the next step, perturbatively). Such a line allows us to keep accurately the wave function of relative motion which provides the leading contribution to the bremsstrahlung spectrum,
while many-nucleon structure should be estimated after
(such a supposition we made from good agreement between theory and experiment for $\alpha$ decay
which was obtained without nucleon structure
in~\cite{Maydanyuk.2003.PTP,Maydanyuk.2006.EPJA,Maydanyuk.2008.EPJA,Giardina.2008.MPLA,%
Maydanyuk.2009.TONPPJ,Maydanyuk.2009.JPS,Maydanyuk.2009.NPA}).

The nonrelativistic Hamiltonian of the system composed from $A$ nucleons with
two-nucleon interactions has this form:
\begin{equation}
  \hat{H} =
  \hat{T} - \hat{T}_{0} +
  \displaystyle\sum\limits_{i>j=1}^{A} \hat{V}(ij) +
  \displaystyle\sum\limits_{i>j=1}^{Z} \displaystyle\frac{e^{2}}{|\mathbf{r}_{i} - \mathbf{r}_{j}|},
\label{eq.2.3.3}
\end{equation}
where
$\hat{T}$ is the operator of kinetic energy of all nucleons in the laboratory frame,
$\hat{T}_{0}$ is the operator of kinetic energy of the center of mass of a system composed of all nucleons in the laboratory frame,
the term with the first summation describes nuclear interactions between two nucleons, the term with the second summation --- two-nucleons Coulomb interactions.
We shall use the many-nucleon function of the nuclear system using
the basics of the algebraic model of the resonating group method%
\footnote{For example, see
\cite{Steshenko.1971.SJNP} for basics of the model of the deformed oscillating shells,
\cite{Filippov.1980.SJNP,Filippov.1981.SJNP,Vasilevsky.1989.SJNP,Vasilevsky.1990.SJNP,Filippov.1985.SJPN} for basics of the model for the binary cluster configurations for light nuclei,
\cite{Filippov.1985.SJPN,Filippov.1986.SJNP,Filippov.1984.NPA,Filippov.1984.SJNP} for its extensions to describe binary clusters coupled to collective (quadrupole and monopole) channels,
\cite{Arickx.1996.conf,Vasilevsky.1997.PAN,Filippov.1994.PPN,Vasilevsky.2001.PRC,Vasilevsky.2012.PRC} for three-cluster configurations
considered in the framework of such a model.}
 in the form of the Slater determinant:
\begin{equation}
\begin{array}{lcl}
  \psi_{\rm nucl, s}(\beta_{A}) \equiv
  \psi_{\rm nucl, s} (1 \ldots A ) =
  \displaystyle\frac{1}{\sqrt{A!}}
  \displaystyle\sum\limits_{p}
    (-1)^{\varepsilon_{p}}
    \psi_{\lambda_{1}}(1)
    \psi_{\lambda_{2}}(2) \ldots
    \psi_{\lambda_{A}}(A).
\end{array}
\label{eq.2.3.4}
\end{equation}
Here a summation is performed over all $A!$ permutations of coordinates or states of nucleons,
$\varepsilon_{p}$ is the number of permutation in the formalism of the determinant wave functions.
One-nucleon functions
$\psi_{\lambda_{s}} (s) = \varphi_{n_{s}} (\mathbf{r}_{s})\, \bigl|\, \sigma^{(s)} \tau^{(s)} \bigr\rangle$
represent the multiplication of space and spin-isospin functions,
where
$\varphi_{n_{s}}$ is a space function of the $s$-th nucleon, 
$n_{s}$ is the number of state of the space function of the $s$-th nucleon, 
$\bigl|\, \sigma^{(s)} \tau^{(s)} \bigr\rangle$ is the corresponding spin-isospin function. 
%
%
We shall study the emission of photons as a perturbation of the nuclear system, defined by the operator of emission $\hat{H}_{\gamma}$. For a description of the emission of photons we shall calculate the matrix element
written via a combination of one-nucleon wave functions in the following form:
\begin{equation}
\begin{array}{lcl}
  \langle \psi_{\rm nucl, f} (1 \ldots A )\, |\, \hat{H}_{\gamma}\, |\, \psi_{\rm nucl, i} (1 \ldots A ) \rangle = \\
  = \quad
    \displaystyle\frac{1}{A\,(A-1)}\;
    \displaystyle\sum\limits_{k=1}^{A}
    \displaystyle\sum\limits_{m=1, m \ne k}^{A}
    \biggl\{
      \langle \psi_{k}(i)\, \psi_{m}(j) |\, \hat{H}_{\gamma}\, |\, \psi_{k}(i)\, \psi_{m}(j) \rangle -
      \langle \psi_{k}(i)\, \psi_{m}(j) |\, \hat{H}_{\gamma}\, |\, \psi_{m}(i)\, \psi_{k}(j) \rangle
    \biggr\}.
\end{array}
\label{eq.2.3.6}
\end{equation}

\subsection{Matrix element of emission
\label{sec.2.4}}

We find the matrix element on the basis of the operator of emission (\ref{eq.2.2.2}) and
wave function (\ref{eq.2.3.1}):
\begin{equation}
\begin{array}{lcl}
  \vspace{0mm}
  \langle \Psi_{f} |\, \hat{H}_{\gamma} |\, \Psi{i} \rangle & = &

  -\, N_{i}\, N_{f}\; e\;
  \sqrt{\displaystyle\frac{2\pi\hbar}{w_{\rm ph}}}\,

    \displaystyle\sum\limits_{\alpha=1,2} \mathbf{e}^{(\alpha),*} \cdot
  \Biggl\{ \\

  \vspace{2mm}
  & & \times \;
  \Biggl\langle \Psi_{f} \Biggl|\,
    e^{i\,(\mathbf{K}_{i} - \mathbf{K}_{f} - \mathbf{k})\cdot\mathbf{R}}\:
    e^{i\, \mathbf{k \cdot r}\, \displaystyle\frac{m_{\rm p}}{m_{A}+m_{\rm p}}}\;
    \displaystyle\frac{1}{m_{A} + m_{\rm p}}\;
    \biggl[
      e^{-i \mathbf{k} \cdot \mathbf{r}}\, z_{\rm p} +
      \displaystyle\sum\limits_{j=1}^{A} z_{Aj}\; e^{-i \mathbf{k} \cdot \rhobfsm_{Aj}}
    \biggr]\; \mathbf{P}\;
    \Biggr|\, \Psi_{i} \Biggr\rangle\; + \\

  \vspace{2mm}
  & & +\;
    \Biggl\langle \Psi_{f} \Biggl|\,
      e^{i\,(\mathbf{K}_{i} - \mathbf{K}_{f} - \mathbf{k})\cdot\mathbf{R}}\:
      e^{i\, \mathbf{k \cdot r}\, \displaystyle\frac{m_{\rm p}}{m_{A}+m_{\rm p}}}\,
    \biggl[
      e^{-i \mathbf{k} \cdot \mathbf{r}}\, \displaystyle\frac{z_{\rm p}}{m_{\rm p}} -
      \displaystyle\sum\limits_{j=1}^{A}
        \displaystyle\frac{z_{Aj}}{m_{A}}\; e^{-i \mathbf{k} \cdot \rhobfsm_{Aj}}
    \biggr]\; \mathbf{p}\;
    \Biggr|\, \Psi_{i} \Biggr\rangle\; + \\

  \vspace{2mm}
  & & +\;
    \Biggl\langle \Psi_{f} \Biggl|\,
      e^{i\,(\mathbf{K}_{i} - \mathbf{K}_{f} - \mathbf{k})\cdot\mathbf{R}}\:
      e^{i\, \mathbf{k \cdot r}\, \displaystyle\frac{m_{\rm p}}{m_{A}+m_{\rm p}}}\,
      \biggl[
        \displaystyle\sum\limits_{j=1}^{A-1}
          \displaystyle\frac{z_{A j}}{m_{Aj}}\:
          e^{-i \mathbf{k} \cdot \rhobfsm_{Aj}}\,
          \mathbf{\tilde{p}}_{Aj}
      \biggr]\; \Biggr|\, \Psi_{i} \Biggr\rangle\; - \\

  \vspace{2mm}
  & & -\;
    \Biggl\langle \Psi_{f} \Biggl|\,
      e^{i\,(\mathbf{K}_{i} - \mathbf{K}_{f} - \mathbf{k})\cdot\mathbf{R}}\:
      e^{i\, \mathbf{k \cdot r}\, \displaystyle\frac{m_{\rm p}}{M+m_{\rm p}}}\,
    \displaystyle\frac{1}{m_{A}}\,
    \biggl[
      \displaystyle\sum\limits_{j=1}^{A}
      z_{Aj}\; e^{-i \mathbf{k} \cdot \rhobfsm_{Aj}}\;
      \displaystyle\sum_{k=1}^{A-1} \mathbf{\tilde{p}}_{Ak}
    \biggr]\; \Biggr|\, \Psi_{i} \Biggr\rangle\;
  \Biggr\}.
\end{array}
\label{eq.2.4.1}
\end{equation}
The first term the describes emission of the photon caused by the motion of the full nuclear system in the laboratory frame and its response on the emission of the photon. We shall calculate the spectra in the center-of-mass frame, and thus shall neglect this term.
The second term describes the emission of the photon caused by a proton and each nucleon of the nucleus, at relative motion of the proton concerning the nucleus. This term contributes the most strongly to the full bremsstrahlung spectrum.
The third and fourth terms describe the emission of the photon caused by each nucleon of the nucleus, in relative motions of nucleons of the nucleus inside its space region
(any nuclear deformations during emission can be connected with such terms).

We shall start from a consideration of the leading matrix element on the basis of the second term
in Eq.~(\ref{eq.2.4.1}) (we shall denote it by bottom index 1).
We integrate over all independent space variables given in Eq.¬(\ref{eq.2.1.3}) and obtain
\begin{equation}
\begin{array}{lcl}
  \langle \Psi_{f} |\, \hat{H}_{\gamma} | \Psi_{i} \rangle_{1} =
  - N_{i} N_{f}\,
  \displaystyle\frac{e}{m}\:
  \sqrt{\displaystyle\frac{2\pi\hbar}{w_{\rm ph}}}\,
  (2\pi)^{3}
    \displaystyle\sum\limits_{\alpha=1,2} \mathbf{e}^{(\alpha),*}
    \delta(\mathbf{K}_{f} - \mathbf{K}_{i} - \mathbf{k}) \cdot
    \biggl\langle \Phi_{\rm p-nucl, f} (\mathbf{r})\, \biggl|\,
      Z_{\rm eff}(\mathbf{k}, \mathbf{r})\: e^{-i\,\mathbf{k \cdot r}}\: \mathbf{p}\,
    \biggr|\, \Phi_{\rm p-nucl, i} (\mathbf{r}) \biggr\rangle,
\end{array}
\label{eq.2.4.5}
\end{equation}
where we have introduced the \emph{effective charge of the proton-nucleus system} as
\begin{equation}
\begin{array}{lcl}
  Z_{\rm eff}(\mathbf{k}, \mathbf{r}) & = &
    e^{i\,\mathbf{k \cdot r}\, \displaystyle\frac{m_{\rm p}}{m_{A}+m_{\rm p}}}\:
    \biggl\{
      \displaystyle\frac{m_{A}\, z_{\rm p}}{m_{A}+m_{\rm p}} -
      e^{i\,\mathbf{k \cdot r}}\:
      \displaystyle\frac{m_{\rm p}\, Z_{\rm A}(\mathbf{k})}{m_{A}+m_{\rm p}}
    \biggr\}
\end{array}
\label{eq.2.4.3}
\end{equation}
and the \definition{charged form factor of the nucleus} as
\begin{equation}
  Z_{\rm A} (\mathbf{k}) =
  \Bigl\langle \psi_{\rm nucl, f} (\beta_{A})\: \Bigl|\; 
    \displaystyle\sum\limits_{j=1}^{A}
      z_{Aj}\:
      e^{-i \mathbf{k} \cdot \rhobfsm_{Aj}}\:
  \Bigr|\: \psi_{\rm nucl, i} (\beta_{A})\, \Bigr\rangle. 
\label{eq.2.4.4}
\end{equation}
Here $m = m_{\rm p} m_{A} / (m_{\rm p} + m_{A})$ is the reduced mass and
we use the integral representation of the $\delta$ function.
We define the normalizing factors $N_{i}$ and $N_{f}$ as $ N_{i} = N_{f} = (2\pi)^{-3/2}$.
We shall calculate cross sections of the emitted photons not dependent on momentum $\mathbf{K}_{f}$ (momentum of the full proton-nucleus system after the emission of a photon in the laboratory frame). So, we have to integrate the matrix element (\ref{eq.2.4.5}) over momentum $\mathbf{K}_{f}$
and we obtain
\begin{equation}
\begin{array}{ll}
  \langle \Psi_{f} |\, \hat{H}_{\gamma} |\, \Psi_{i} \rangle_{1} =
  -\,\displaystyle\frac{e}{m}\:
  \sqrt{\displaystyle\frac{2\pi\hbar}{w_{\rm ph}}}\;
    \displaystyle\sum\limits_{\alpha=1,2} \mathbf{e}^{(\alpha),*} \cdot
    \biggl\langle\: \Phi_{\rm p-nucl, f} (\mathbf{r})\: \biggl|\,
      Z_{\rm eff}(\mathbf{k}, \mathbf{r})\: e^{-i\,\mathbf{k \cdot r}}\: \mathbf{p}\;
    \biggr|\: \Phi_{\rm p-nucl, i} (\mathbf{r})\: \biggr\rangle, &
  \mathbf{K}_{i} = \mathbf{K}_{f} + \mathbf{k}.
\end{array}
\label{eq.2.4.8}
\end{equation}
The effective charge of the system in the first approximation $\exp(i\, \mathbf{k \cdot r}) \to 1$ (called \definition{dipole} concerning the effective charge)
obtains the form
\begin{equation}
  Z_{\rm eff}^{\rm (dip)} (\mathbf{k}) =
  \displaystyle\frac{m_{A}\, z_{\rm p} - m_{\rm p}\, Z_{\rm A}(\mathbf{k})}{m_{A}+m_{\rm p}}.
\label{eq.2.4.10}
\end{equation}
It is apparent that in such an approximation the effective charge is independent on the relative distance between the proton and center of mass of the nucleus.

The simplest matrix element is obtained by neglecting relative displacements of nucleons of the nucleus inside its space region (i.e. in approximation where the nucleus is considered as point-like and we use $e^{-i \mathbf{k} \cdot \rhobfsm_{Aj}} \to 1$ for each nucleon).
The form factor of the nucleus represents the summarized electromagnetic charge of the nucleons of the nucleus,
$Z_{\rm A} (\mathbf{k}) \to
Z_{\rm A}$,
where the dependence on characteristics of the emitted photon is lost as the functions $\psi_{\rm nucl, s}$ are normalized
(see Appendix~\ref{sec.app.1} for details).
At such approximations we obtain the matrix element [we add the upper index $\rm (dip)$]:
\begin{equation}
\begin{array}{lcl}
\vspace{2mm}
  \langle \Psi_{f} |\, \hat{H}_{\gamma} |\, \Psi_{i} \rangle_{1}^{\rm (dip)} \;\; = \;\;
  \displaystyle\frac{e}{m}\:
    \sqrt{\displaystyle\frac{2\pi\hbar}{w_{\rm ph}}}\:
    p_{1}\: 2\pi\; \delta(w_{i} - w_{f} - w), &

  Z_{\rm eff}^{\rm (dip,0)} = \displaystyle\frac{m_{A}\, z_{\rm p} - m_{\rm p}\, Z_{\rm A}}{m_{A}+m_{\rm p}}, \\

  p_{1} \;\; = \;\;
  -\, Z_{\rm eff}^{\rm (dip,0)}\,
    \displaystyle\sum\limits_{\alpha=1,2} \mathbf{e}^{(\alpha),*} \cdot
    \Bigl\langle\: \psi_{\rm p-nucl, f} (\mathbf{r})\: \Bigl|\,
      e^{-i\,\mathbf{k \cdot r}}\: \mathbf{p}\;
    \Bigr|\: \psi_{\rm p-nucl, i} (\mathbf{r})\: \Bigr\rangle,
\end{array}
\label{eq.2.4.14}
\end{equation}
where the wave packets
\begin{equation}
  \Phi_{\rm p-nucl, s} (\mathbf{r}, t) =
  \displaystyle\int\limits_{0}^{+\infty}
    g(k - k_{s})\:
    \psi_{\rm p-nucl, s} (\mathbf{r})\:
    e^{-iw(k)t}\; dk
\label{eq.2.4.13}
\end{equation}
are used as the functions $\psi_{\rm p-nucl, s} (\mathbf{r})$ (as in the formalism of \cite{Maydanyuk.2003.PTP}).
Such a matrix element $p_{1}$ coincides exactly with the electrical matrix element $p_{\rm el}$ in Eq.~(10) in the dipole approximation of the effective charge in~\cite{Maydanyuk.2012.PRC} without the inclusion of spin states of the scattered proton.


\subsection{Emission formed by displacements of nucleons inside the nucleus
\label{sec.2.5}}

Now we shall find the correction to the matrix element (\ref{eq.2.4.14}) taking into account displacements of nucleons of the nucleus inside its space region
(we shall denote such correction by bottom index 2).
Thus, we write the matrix element (\ref{eq.2.4.8}) as
\begin{equation}
\begin{array}{lcl}
  \vspace{0mm}
  \langle \Psi_{f} |\, \hat{H}_{\gamma} |\, \Psi_{i} \rangle_{1} & = &
  \langle \Psi_{f} |\, \hat{H}_{\gamma} |\, \Psi_{i} \rangle_{1}^{\rm (dip)} +
  \langle \Psi_{f} |\, \hat{H}_{\gamma} |\, \Psi_{i} \rangle_{2}
\end{array}
\label{eq.2.5.1}
\end{equation}
and find the correction
[after use of wave functions $\Phi_{\rm p-nucl,\: s} (\mathbf{r})$ as Eq.~(\ref{eq.2.4.13})]:
\begin{equation}
\begin{array}{lcllcl}
\vspace{1mm}
  \langle \Psi_{f} |\, \hat{H}_{\gamma} |\, \Psi_{i} \rangle_{2} \;\; = \;\;
    \displaystyle\frac{e}{m}\: \sqrt{\displaystyle\frac{2\pi\hbar}{w_{\rm ph}}}\: p_{2}\:
    2\pi\; \delta(w_{i} - w_{f} - w), \\

  p_{2} \;\; = \;\;
  - \displaystyle\sum\limits_{\alpha=1,2} \mathbf{e}^{(\alpha),*} \cdot
  \biggl\langle \psi_{\rm p-nucl, f} (\mathbf{r})\: \biggl|\;
    Z_{\rm eff}^{(2)} (\mathbf{k}, \mathbf{r})\, e^{-i\,\mathbf{k \cdot r}}\: \mathbf{p}\;
  \biggr|\: \psi_{\rm p-nucl, i} (\mathbf{r}) \biggr\rangle,
\end{array}
\label{eq.2.5.2}
\end{equation}
where a new correction for the effective charge is introduced in the form
\begin{equation}
\begin{array}{lcl}
  Z_{\rm eff}^{(2)} (\mathbf{k}, \mathbf{r}) & = &
  \biggl( e^{i\,\mathbf{k \cdot r}\, \displaystyle\frac{m_{\rm p}}{m_{A}+m_{\rm p}}} - 1 \biggr)\,
    \displaystyle\frac{m_{A}\, z_{\rm p}}{m_{A}+m_{\rm p}}\; -\;
    \displaystyle\frac{m_{\rm p}}{m_{A}+m_{\rm p}}\;
  \Bigl\{
    e^{i\,\mathbf{k \cdot r}\, \displaystyle\frac{m_{\rm p}}{m_{A}+m_{\rm p}}}\, e^{i\,\mathbf{k \cdot r}}\,
    Z_{A}(\mathbf{k}) - Z_{A}
  \Bigr\}.
\end{array}
\label{eq.2.5.5}
\end{equation}
In the dipole approximation for the effective charge we have
[see Appendix~\ref{sec.app.1} for calculations of $Z_{A}(\mathbf{k})$]:
\begin{equation}
\begin{array}{lcl}
  Z_{\rm eff}^{\rm (dip,\, 2)} (\mathbf{k}) & = &
  -\: \displaystyle\frac{m_{\rm p}}{m_{A}+m_{\rm p}}\:
  \Bigl( Z_{\rm A} (\mathbf{k}) - Z_{\rm A} \Bigr).
\end{array}
\label{eq.2.5.6}
\end{equation}
One can see that such a function gives correction to the electromagnetic charge of the nucleus.
As an exponential factor in the matrix element (\ref{eq.2.5.2}) has less unity, the correction to the charge of the nucleus is less than this charge (that explains it as a correction to the charge).
In general, the correction reduces the total charge of the nucleus (as a result of a not point-like space consideration of the nucleus).
As we consider nucleons of the nucleus in the bound states, the matrix element should be calculated without divergencies.

The matrix element constructed on the basis of the dipole effective charge (\ref{eq.2.5.6}) has a more simple form:
\begin{equation}
\begin{array}{lcl}
  p_{2}^{\rm (dip)} & = &
  -\,
  Z_{\rm eff}^{\rm (dip,\, 2)} (\mathbf{k})\;
  \displaystyle\sum\limits_{\alpha=1,2} \mathbf{e}^{(\alpha),*} \cdot
  \Bigl\langle \psi_{\rm p-nucl, f} (\mathbf{r})\: \Bigl|\;
    e^{-i\,\mathbf{k \cdot r}}\: \mathbf{p}\;
  \Bigr|\: \psi_{\rm p-nucl, i} (\mathbf{r}) \Bigr\rangle.
\end{array}
\label{eq.2.5.7}
\end{equation}
%
If we wish to include parameters of the emitted photons into the nuclear form factor, we must calculate the matrix element outside the dipole approximation. In such a case, one can use the formula (\ref{eq.2.5.2})
where the more accurate representation of the effective charge is
(see Appendix~\ref{sec.app.2}, for details)
\begin{equation}
\begin{array}{lcl}
  Z_{\rm eff}^{(2)} (\mathbf{k}, \mathbf{r}) =
  \displaystyle\sum\limits_{l=0}^{+\infty}
    i^{l}\, (2l+1)\, P_{l} (\cos \beta)\:
    Z_{\rm eff,\, l}^{(2)} (\mathbf{k}, r)\; - \;
  Z_{\rm eff}^{\rm (dip, 0)},
\end{array}
\label{eq.2.6.5}
\end{equation}
where \emph{partial components of the effective charge} are introduced as
\begin{equation}
\begin{array}{lcl}
  Z_{\rm eff,\, l}^{(2)} (\mathbf{k}, r) & = &
  \displaystyle\frac{m_{A}\, z_{\rm p}}{m_{A}+m_{\rm p}}\;
    j_{l} \Bigl(\displaystyle\frac{m_{\rm p}}{m_{A}+m_{\rm p}}\, kr \Bigr)\; -
  \displaystyle\frac{m_{\rm p}\, Z_{A}(\mathbf{k})}{m_{A}+m_{\rm p}}\,
    j_{l} \Bigl( \displaystyle\frac{m_{A} + 2\,m_{\rm p}}{m_{A}+m_{\rm p}}\, kr \Bigr)
\end{array}
\label{eq.2.6.6}
\end{equation}
and $\beta$ is angle between vectors $\mathbf{k}$ and $\mathbf{r}$.
From such a formula one can see that on smaller distances (of variable $r$) the first term should be dominated in the integration of the matrix element, but on far distances the second term (which is decreased more slowly) has a larger contribution.
Such an effective charge should change the shape of the bremsstrahlung spectrum as it changes the dependence of the matrix element on the energy of photon.


Now we shall consider the emission of photons determined by the third matrix element in Eq.~(\ref{eq.2.4.1}) (we shall denote such a matrix element by bottom index \emph{add}).
Performing an integration over space variable $\mathbf{R}$, momentum $\mathbf{K}$, using the normalizing factors $N_{i}$ and $N_{f}$
and the packets (\ref{eq.2.4.13}) as functions $\Psi_{\rm p-nucl,\: s} (\mathbf{r})$,
we obtain (see Appendix~\ref{sec.app.2} for details)
\begin{equation}
\begin{array}{ccl}
  \langle \Psi_{f} |\, \hat{H}_{\gamma} |\, \Psi_{i} \rangle_{\rm add} =
  \displaystyle\frac{e}{m}\:
    \sqrt{\displaystyle\frac{2\pi\hbar}{w_{\rm ph}}}\; p_{fi,\: \rm add}\;
    2\pi\; \delta(w_{i} - w_{f} - w), &

  \mathbf{K}_{i} = \mathbf{K}_{f} + \mathbf{k},
\end{array}
\label{eq.2.7.2}
\end{equation}
where
\begin{equation}
\begin{array}{lcl}
  p_{fi,\: \rm add} & = &
  -\,\mu\;
  \displaystyle\sum\limits_{l=0}^{+\infty}
    i^{l}\, (2l+1)\, P_{l} (\cos \beta)\; M_{l} (k) \,
  \displaystyle\sum\limits_{\alpha=1,2} \mathbf{e}^{(\alpha),*} \cdot
    \mathbf{D}_{\rm A} (\mathbf{k}),
\end{array}
\label{eq.2.7.5}
\end{equation}
\begin{equation}
\begin{array}{lcl}
  \mathbf{D}_{\rm A} (\mathbf{k}) & = &
  \Bigl\langle \psi_{\rm nucl, f} (\beta_{A})\: \Bigl|\,
    \displaystyle\sum\limits_{j=1}^{A-1}
      \displaystyle\frac{z_{A j}}{m_{Aj}}\:
      e^{-i\, \mathbf{k} \cdot \rhobfsm_{Aj}}\,
      \mathbf{\tilde{p}}_{Aj}
  \Bigr|\, \psi_{\rm nucl, i} (\beta_{A})\: \Bigr\rangle,
\end{array}
\label{eq.2.7.3}
\end{equation}
and we have introduced the nucleon partial matrix elements as
\begin{equation}
\begin{array}{lcl}
  M_{l} (k) & = &
  \biggl\langle \psi_{\rm p-nucl, f} (\mathbf{r})\: \biggl|\,
    j_{l} \Bigl(\displaystyle\frac{m_{\rm p}}{m_{A}+m_{\rm p}} kr \Bigr)\,
  \biggr|\, \psi_{\rm p-nucl, i} (\mathbf{r})\: \biggr\rangle.
\end{array}
\label{eq.2.7.6}
\end{equation}
Taking into account Coulomb gauge and solution (\ref{eq.app.3.10}) for the function $\mathbf{D}_{\rm A} (\mathbf{k})$ given in Appendix~\ref{sec.app.3}, we find that the matrix element (\ref{eq.2.7.5}) equals zero. By the same reason, the last matrix element in Eq.~(\ref{eq.2.4.1}) equals zero also.


\subsection{Inclusion of spin states of the scattering proton
\label{sec.2.8}}

The operator of emission of photons in spinor formalism of the scattering proton,
has the following form~\cite{Maydanyuk.2012.PRC}:
\begin{equation}
\begin{array}{lcl}
  \hat{H}_{\gamma} & = &
  Z_{\rm eff}\, \displaystyle\frac{e}{m c}\,
  \sqrt{\displaystyle\frac{2\pi\hbar c^{2}}{w_{\rm ph}}}\;
    \displaystyle\sum\limits_{\alpha=1,2}
  e^{-i\,\mathbf{k_{\rm ph} \cdot r}}\;
  \Big(
    i\, \mathbf{e}^{(\alpha)}\, \nabla -
    \displaystyle\frac{1}{2}\: \sigmabf\cdot
      \Bigl[ \nabla \times \mathbf{e}^{(\alpha)} \Bigr] +
    i\,\displaystyle\frac{1}{2}\: \sigmabf\cdot
      \Bigl[ \mathbf{k}_{\rm ph} \times \mathbf{e}^{(\alpha)} \Bigr]
  \Bigr).
\end{array}
\label{eq.2.8.1}
\end{equation}
Now the stationary wave function of the scattering proton [i.e., the function $\psi_{\rm p-nucl, s} (\mathbf{r})$ above] is in the form of a bilinear combination of eigenfunctions of orbital and spin subsystems
(see also eq.~(1.4.2) in \cite{Ahiezer.1981}, p.~42),
which were studied in details in~\cite{Maydanyuk.2012.PRC}.
However, we assume that it is not possible experimentally to fix states for selected $M$
(eigenvalue of momentum operator $\hat{J}_{z}$).
So, we shall be interesting in our superposition over all states with different $M$ and
define the wave function so
\begin{equation}
  \psi_{{\rm p-nucl},\: jl} (\mathbf{r}, s) =
  R\,(r)\:
  \displaystyle\sum\limits_{m=-l}^{l}
  \displaystyle\sum\limits_{\mu = \pm 1/2}
    C_{lm 1/2 \mu}^{j,M=m+\mu}\, Y_{lm}(\mathbf{n_{r}})\, v_{\mu} (s),
\label{eq.2.8.2}
\end{equation}
where $R\,(r)$ is radial scalar function (not dependent on different $m$ at the same $l$),
$\mathbf{n}_{\rm r} = \mathbf{r} / r$ is unit vector directed along $\mathbf{r}$,
$Y_{lm}(\mathbf{n}_{\rm r})$ are spherical functions (we use definition (28,7)--(28,8), p.~119
in~\cite{Landau.v3.1989}),
$C_{lm 1/2 \mu}^{jM}$ are Clebsch-Gordon coefficients, $s$ is a variable of spin,
$M = m + \mu$ and $l = j \pm 1/2$.
For convenience, we introduce the space wave function in the form of
$\varphi_{{\rm p-nucl,}\: lm} (\mathbf{r}) = R_{l}\,(r)\: Y_{lm}(\mathbf{n_{\rm r}})$.

So, after the inclusion of the spin formalism of the scattering proton (see eqs.~(10) and (36) in~\cite{Maydanyuk.2012.PRC}),
we obtain the updated formulas (\ref{eq.2.4.14}), (\ref{eq.2.5.7}) and formula (\ref{eq.2.5.2}) with
effective charge (\ref{eq.2.6.5}) as
\begin{equation}
\begin{array}{ll}
  p_{1} =
  Z_{\rm eff}^{\rm (dip,0)}
  \sqrt{\displaystyle\frac{\pi}{2}}
  \displaystyle\sum\limits_{l_{\rm ph}=1}
    (-i)^{l_{\rm ph}} \sqrt{2l_{\rm ph}+1}
  \displaystyle\sum\limits_{m_{i}, m_{f}}
  \displaystyle\sum\limits_{\mu_{i}, \mu_{f} = \pm 1/2}
    \hspace{-3mm}
    C_{l_{f}m_{f} 1/2 \mu_{f}}^{j_{f} M_{f}=m_{f}+\mu_{f},\,*}
    C_{l_{i}m_{i} 1/2 \mu_{i}}^{j_{i}M_{i}=m_{i}+\mu_{i}}
  \displaystyle\sum\limits_{\mu=\pm 1}
    h_{\mu}
    \Bigl[
      i\mu p_{l_{\rm ph}\mu}^{M m_{i} m_{f}} +
      p_{l_{\rm ph}\mu}^{E m_{i} m_{f}}
    \Bigr],
\end{array}
\label{eq.2.8.4}
\end{equation}
\begin{equation}
\begin{array}{ll}
  p_{2}^{\rm (dip)} =
  Z_{\rm eff}^{\rm (dip,2)} (\mathbf{k})
  \sqrt{\displaystyle\frac{\pi}{2}}\,
  \hspace{-0.5mm}
  \displaystyle\sum\limits_{l_{\rm ph}=1}
    \hspace{-0.5mm}
    (-i)^{l_{\rm ph}} \sqrt{2l_{\rm ph}+1}
  \hspace{-0.8mm}
  \displaystyle\sum\limits_{m_{i}, m_{f}}
  \displaystyle\sum\limits_{\mu_{i}, \mu_{f} = \pm 1/2}
    \hspace{-4.5mm}
    C_{l_{f}m_{f} 1/2 \mu_{f}}^{j_{f} M_{f}=m_{f}+\mu_{f},\,*}
    C_{l_{i}m_{i} 1/2 \mu_{i}}^{j_{i}M_{i}=m_{i}+\mu_{i}}
  \hspace{-2mm}
  \displaystyle\sum\limits_{\mu=\pm 1}
    \hspace{-1.5mm} h_{\mu}
    \Bigl[
      i\mu p_{l_{\rm ph}\mu}^{M m_{i} m_{f}}
      \hspace{-0.7mm} + \hspace{-0.5mm}
      p_{l_{\rm ph}\mu}^{E m_{i} m_{f}}
    \Bigr],
\end{array}
\label{eq.2.8.5}
\end{equation}
\begin{equation}
\begin{array}{ll}
  p_{3} =
  \sqrt{\displaystyle\frac{\pi}{2}}
  \displaystyle\sum\limits_{l_{\rm ph}=1}
    (-i)^{l_{\rm ph}} \sqrt{2l_{\rm ph}+1}
  \displaystyle\sum\limits_{m_{i}, m_{f}}
  \displaystyle\sum\limits_{\mu_{i},\, \mu_{f} = \pm 1/2}
    \hspace{-3mm}
    C_{l_{f}m_{f} 1/2 \mu_{f}}^{j_{f} M_{f}=m_{f}+\mu_{f},\,*}\,
    C_{l_{i}m_{i} 1/2 \mu_{i}}^{j_{i}M_{i}=m_{i}+\mu_{i}}
  \displaystyle\sum\limits_{\mu=\pm 1}
    h_{\mu}
    \Bigl[
      i\mu \breve{p}_{l_{\rm ph}\mu}^{M m_{i} m_{f}} +
      \breve{p}_{l_{\rm ph}\mu}^{E m_{i} m_{f}}
    \Bigr],
\end{array}
\label{eq.2.8.6}
\end{equation}
The matrix elements
$p_{l_{\rm ph}\mu}^{M m_{i} m_{f}}$,
$p_{l_{\rm ph}\mu}^{E m_{i} m_{f}}$,
$\breve{p}_{l_{\rm ph}\mu}^{M m_{i} m_{f}}$ and
$\breve{p}_{l_{\rm ph}\mu}^{E m_{i} m_{f}}$
are given in Appendix~\ref{sec.app.4}.

\subsection{Correction to the emission of photons from relative momenta of nucleons inside the nucleus,
caused by taking spin states of the scattering proton into account
\label{sec.2.9}}

The strongest emission is formed by the first term in the emission operator (\ref{eq.2.8.2}), which we formulated in the sections above. The next by intensity emission is formed by the second term in Eq.~(\ref{eq.2.8.2}) (according to analysis in~\cite{Maydanyuk.2012.PRC}), which we shall study in this subsection. The last term in Eq.~(\ref{eq.2.8.2}) gives the smallest emission which will be neglected in this paper.
The corresponding matrix element with the included second term of the emission operator,
after integration over space variable $\mathbf{R}$ and momentum $\mathbf{K}$,
using packets (\ref{eq.2.4.13}) for wave function $\Phi_{\rm p-nucl,\: s}(\mathbf{r})$ of the scattering proton,
is (we shall denote it by bottom index 4)
\begin{equation}
\begin{array}{cc}
  \langle \Psi_{f} |\, \hat{H}_{\gamma} |\, \Psi_{i} \rangle_{4} =
  \displaystyle\frac{e}{m}\; \sqrt{\displaystyle\frac{2\pi\hbar}{w_{\rm ph}}}\; p_{4}\; 2\pi\; \delta(w_{i} - w_{f} - w),
\end{array}
\label{eq.2.9.1}
\end{equation}
where
\begin{equation}
\begin{array}{lcl}
\vspace{0.7mm}
  p_{4} & = &
  m\: \displaystyle\sum\limits_{\alpha=1,2}
    \Biggl\langle
      \psi_{\rm p-nucl, f}(\mathbf{r}) \cdot
      \psi_{\rm nucl, f}(\beta_{A})
    \Biggl|\,
      e^{i\, \mathbf{k \cdot r}\, \displaystyle\frac{m_{\rm p}}{m_{A}+m_{\rm p}}}\; \times \\
& \times &
      \biggl\{
        \displaystyle\sum\limits_{j=1}^{A-1}
          \displaystyle\frac{z_{A j}}{m_{Aj}}\:
          e^{-i \mathbf{k} \cdot \rhobfsm_{Aj}}
          \displaystyle\frac{1}{2}\: \sigmabf \cdot
            \Bigl[ \mathbf{\tilde{p}}_{Aj} \times \mathbf{e}^{(\alpha),*} \Bigr]
      \biggr\}\:
    \Biggl|\,
      \psi_{\rm p-nucl, i}(\mathbf{r}) \cdot
      \psi_{\rm nucl, i}(\beta_{A})
    \Biggr\rangle.
\end{array}
\label{eq.2.9.2}
\end{equation}
This matrix element can be separated on two integrals.
Using solution (\ref{eq.app.3.10}) for the function $\mathbf{D}_{\rm A} (\mathbf{k})$ and summarizing over polarization states of the emitted photon [using property (\ref{eq.2.2.3})],
we obtain
\begin{equation}
\begin{array}{lcl}
  p_{4} & = &
  \displaystyle\frac{\hbar\,m\, k}{4}\:
  Z_{\rm A} (\mathbf{k})\,
  \biggl\langle \psi_{\rm p-nucl, f} (\mathbf{r})\: \biggl|\,
    e^{i\, \mathbf{k \cdot r}\, \displaystyle\frac{m_{\rm p}}{m_{A}+m_{\rm p}}}\, \sigmabf\:
  \biggr|\, \psi_{\rm p-nucl, i} (\mathbf{r}) \biggr\rangle\: \cdot\:
  \bigl(\mathbf{e}^{(2),*} - \mathbf{e}^{(1),*}\bigr).
\end{array}
\label{eq.2.9.4}
\end{equation}

For taking into account spin states of the scattering proton, we use wave function $\psi_{\rm p-nucl}$
of this proton in form (\ref{eq.2.8.2}) and obtain the following form for
the matrix element (see Ref.~\cite{Maydanyuk.2012.PRC}, for some details):
\begin{equation}
\begin{array}{lcl}
\vspace{0mm}
  p_{4} & = &
  \displaystyle\frac{\hbar\,m\, k}{4}\:
  Z_{\rm A} (\mathbf{k})\,
  \displaystyle\sum\limits_{m_{f}=-l_{f}}^{l_{f}}
  \displaystyle\sum\limits_{m_{i}=-l_{i}}^{l_{i}}
  \displaystyle\sum\limits_{\mu_{i},\, \mu_{f} = \pm 1/2}
    C_{l_{f}m_{f} 1/2 \mu_{f}}^{j_{f},M=m_{f}+\mu_{f},\: *}\,
    C_{l_{i}m_{i} 1/2 \mu_{i}}^{j_{i},M=m_{i}+\mu_{i}}\; \times \\

  & \times &
  \biggl\langle
    \varphi_{{\rm p-nucl,}\, l_{f}m_{f}} (\mathbf{r})\,
  \biggl|\,
    e^{i\, \mathbf{k \cdot r}\, \displaystyle\frac{m_{\rm p}}{m_{A}+m_{\rm p}}}\,
  \biggr|\,
    \varphi_{{\rm p-nucl,}\, l_{i}m_{i}} (\mathbf{r})
  \biggr\rangle \:
  \Bigl\{ - 1 + i\, \bigl[ \delta_{\mu_{i}, +1/2}\; -\; \delta_{\mu_{i}, -1/2} \bigr] \Bigr\}.
\end{array}
\label{eq.2.9.5}
\end{equation}
%
%
%
%
We apply the multipole expansion for the internal matrix element
(components $\tilde{p}_{l_{\rm ph}\mu}^{M}$, $\tilde{p}_{l_{\rm ph}\mu}^{E}$,
corresponding radial and angular integrals are given in Appendix~\ref{sec.app.4}):
\begin{equation}
  \biggl\langle
    \varphi_{{\rm p-nucl,}\, l_{f}m_{f}} (\mathbf{r})\,
  \biggl|\,
    e^{i\, \mathbf{k \cdot r}\, \displaystyle\frac{m_{\rm p}}{m_{A}+m_{\rm p}}}\,
  \biggr|\,
    \varphi_{{\rm p-nucl,}\, l_{i}m_{i}} (\mathbf{r})
  \biggr\rangle =
  \sqrt{\displaystyle\frac{\pi}{2}}\:
  \displaystyle\sum\limits_{l_{\rm ph}=1}\,
    (-i)^{l_{\rm ph}}\, \sqrt{2l_{\rm ph}+1}\;
  \displaystyle\sum\limits_{\mu = \pm 1}
    \Bigl[ \mu\,\tilde{p}_{l_{\rm ph}\mu}^{M} - i\, \tilde{p}_{l_{\rm ph}\mu}^{E} \Bigr],
\label{eq.2.9.7}
\end{equation}
and the total matrix element (\ref{eq.2.9.5}) obtains the form:
\begin{equation}
\begin{array}{lcl}
\vspace{0mm}
  p_{4} & = &
  \displaystyle\frac{\hbar\,m\, k}{4}\:
  Z_{\rm A} (\mathbf{k}) \:
  \sqrt{\displaystyle\frac{\pi}{2}}\:
  \displaystyle\sum\limits_{l_{\rm ph}=1}\,
    (-i)^{l_{\rm ph}}\, \sqrt{2l_{\rm ph}+1} \:
  \displaystyle\sum\limits_{m_{f}=-l_{f}}^{l_{f}}
  \displaystyle\sum\limits_{m_{i}=-l_{i}}^{l_{i}}
  \displaystyle\sum\limits_{\mu_{i},\, \mu_{f} = \pm 1/2}
    C_{l_{f}m_{f} 1/2 \mu_{f}}^{j_{f},M=m_{f}+\mu_{f},\: *}\,
    C_{l_{i}m_{i} 1/2 \mu_{i}}^{j_{i},M=m_{i}+\mu_{i}}\; \times \\

  & \times &
  \Bigl\{ - 1 + i\, \bigl[ \delta_{\mu_{i}, +1/2}\; -\; \delta_{\mu_{i}, -1/2} \bigr] \Bigr\}\:
  \displaystyle\sum\limits_{\mu = \pm 1}
  \Bigl[ \mu\,\tilde{p}_{l_{\rm ph}\mu}^{M} - i\, \tilde{p}_{l_{\rm ph}\mu}^{E} \Bigr].
\end{array}
\label{eq.2.9.8}
\end{equation}

Now we find the total matrix element of the emitted photons. 
After lengthly calculations, we obtain:
%
\begin{equation}
\begin{array}{lcl}
  p_{1} + p_{3} + p_{4} & = &
  \sqrt{\displaystyle\frac{\pi}{2}}\:
  \displaystyle\sum\limits_{l_{\rm ph}=1}\,
    (-i)^{l_{\rm ph}}\, \sqrt{2l_{\rm ph}+1}\;
  \displaystyle\sum\limits_{\mu=\pm 1}
  \displaystyle\sum\limits_{m_{i}, m_{f}}
  \displaystyle\sum\limits_{\mu_{i},\, \mu_{f} = \pm 1/2}
    C_{l_{f}m_{f} 1/2 \mu_{f}}^{j_{f},M=m_{f}+\mu_{f},\,*}\,
    C_{l_{i}m_{i} 1/2 \mu_{i}}^{j_{i},M=m_{i}+\mu_{i}}\, h_{\mu}\; g,
\end{array}
\label{eq.2.9.9}
\end{equation}
where
\begin{equation}
\begin{array}{lcl}
  g & = &
    i\,\mu\,
      \Bigl(
        Z_{\rm eff}^{\rm (dip,0)}\: p_{l_{\rm ph}\mu}^{M m_{i} m_{f}} +
        \breve{p}_{l_{\rm ph}\mu}^{M m_{i} m_{f}}
      \Bigr)\; +
      \Bigl(
        Z_{\rm eff}^{\rm (dip,0)}\: p_{l_{\rm ph}\mu}^{E m_{i} m_{f}} +
        \breve{p}_{l_{\rm ph}\mu}^{E m_{i} m_{f}}
      \Bigr) +
      f\; \displaystyle\frac{\hbar\,m\, k}{4}\:
      Z_{\rm A} (\mathbf{k})\;
      \Bigl(
        i\,\mu\, \tilde{p}_{l_{\rm ph}\mu}^{M} +
        \tilde{p}_{l_{\rm ph}\mu}^{E}
      \Bigr).
%
\end{array}
\label{eq.2.9.10}
\end{equation}
In order to simply estimate how intensive the emission formed on the basis of the dynamics of the nucleons inside nucleus and determined by the contribution (\ref{eq.2.9.8}) [caused by the second term in the emission operator (\ref{eq.2.8.1})] is, on the background of the full bremsstrahlung emission, we introduce a new factor $f$.
Such an introduction of the unified coefficient
allows us to obtain a clear understanding about such a type of emission,
in order to investigate the role of dynamics of nucleons inside the nucleus in the emission of photons.
%
%
It turns out that the calculated spectra are sensitive to values of this coefficient.
So, comparing the calculations with the experimental data this coefficient can be found,
and it will characterize the real contribution of the emission formed by dynamics of nucleons in nucleus.

\subsection{The bremsstrahlung probability and parameters of the proton-nucleus potential
\label{sec.2.10}}

We define the cross section of the emitted photons on the basis of the matrix element (\ref{eq.2.4.1}) (where we include the operator of emission (\ref{eq.2.8.1}) without the last term which gives the smallest contribution into the total spectrum, see Ref.~\cite{Maydanyuk.2012.PRC} for details)
in the framework of the formalism given in Ref.~\cite{Maydanyuk.2012.PRC}.
%
%
%
We calculate the radial wave functions $R_{l}(r)$ numerically concerning the chosen potential
of the interaction between the proton and the spherically symmetric core.
For a description of the proton-nucleus interaction we use the potential as
$V (r) = v_{c} (r) + v_{N} (r) + v_{\rm so} (r) + v_{l} (r)$,
where $v_{c} (r)$, $v_{N} (r)$, $v_{\rm so} (r)$, and $v_{l} (r)$ are Coulomb,
nuclear, spin-orbital, and centrifugal components
having the form~\cite{Becchetti.1969.PR}
\begin{equation}
\begin{array}{ccc}
  \vspace{1mm}
  v_{N} (r) = - \displaystyle\frac{V_{R}} {1 + \exp{\displaystyle\frac{r-R_{R}} {a_{R}}}},
  \hspace{2mm}
  v_{l} (r) = \displaystyle\frac{l\,(l+1)} {2mr^{2}}, 
  \hspace{2mm}
  v_{\rm so} (r) =
    V_{\rm so}\,
    {\mathbf {q \cdot l}}\,
    \displaystyle\frac{\lambda_{\pi}^{2}}{r}\,
    \displaystyle\frac{d}{dr}\, \Bigl[1
    + \exp\Bigl(\displaystyle\frac{r-R_{\rm so}} {a_{\rm so}} \Bigr)\Bigr]^{-1}, \\
  v_{c} (r) =
  \left\{
  \begin{array}{ll}
    \displaystyle\frac{Z e^{2}} {r}, &
      \mbox{at  } r \ge R_{c}, \\
    \displaystyle\frac{Z e^{2}} {2 R_{c}}\;
      \biggl\{ 3 -  \displaystyle\frac{r^{2}}{R_{c}^{2}} \biggr\}, &
      \mbox{at  } r < R_{c}.
  \end{array}
  \right.
\end{array}
\label{eq.2.11.1}
\end{equation}
We use the parametrization proposed by Becchetti and Greenlees in~\cite{Becchetti.1969.PR}
which has been tested in numerous research papers:
\begin{equation}
\begin{array}{ll}
\begin{array}{llll}
  V_{R} = 54.0 - 0.32\,E + 0.4\,Z/A^{1/3} + 24.0\,I, &
  V_{\rm so} =6.2, \\
\end{array} \\
\begin{array}{llll}
  R_{R} = r_{R}\, A^{1/3}, &
  R_{c} = r_{c}\, A^{1/3}, &
  R_{\rm so} = r_{\rm so}\, A^{1/3}, \\
  r_{\rm so} = 1.01\;{\rm fm}, &
  a_{R} = 0.75\; {\rm fm}, &
  a_{\rm so} = 0.75\; {\rm fm}.
\end{array}
\end{array}
\label{eq.2.11.2}
\end{equation}
Here,
$I = (N-Z)/A$, $A$ and $Z$ are the mass and proton numbers of the daughter nucleus,
$E$ is the incident laboratory energy,
$V_{R}$ and $V_{\rm so}$ are the strength of the nuclear and spin-orbital
components defined in MeV,
$R_{c}$ and $R_{R}$ are Coulomb and nuclear radii of the nucleus, $a_{R}$
and $a_{\rm so}$ are diffusion parameters.
The criterion function of the theoretical fit is taken to be
\begin{equation}
  \varepsilon =
  \displaystyle\frac{1}{n_{\rm max}}
  \displaystyle\sum\limits_{n=1}^{n_{\rm max}}
    \Bigl|\sigma^{\rm (theor)} (E_{n}) - \sigma^{\rm (exp)} (E_{n}) \Bigr|,
\label{eq.2.11.3}
\end{equation}
where $\sigma^{\rm (theor)} (E_{n})$ and $\sigma^{\rm (exp)} (E_{n})$ are the
theoretical and experimental values of the bremsstrahlung cross sections
for the chosen nucleus at energy $E_{n}$, and a summation is performed over
all values of experimental data. We shall look for the value for $r_{R}$
(in the first calculations we shall restrict ourselves by approximation
$r_{c} = r_{R}$) when this error (\ref{eq.2.11.3}) is the minimum
(for simplicity, we shall call such an approach the \emph{method of minimization}).
We found a slower sensitivity of the bremsstrahlung spectra on $V_{R}$ in
comparison with $r_{R}$. So, in this paper we shall study the influence of
the parameter $r_{R}$ on the spectra at fixed $V_{R}$ given
by formula~(\ref{eq.2.11.3}).


\section{Analysis
\label{sec.results}}

Our analysis begins with the $p + ^{208}{\rm Pb}$ reaction, which has been intensively studied by different experimental groups and, so, has proper experimental material~\cite{Edington.1966.NP,Clayton.1991.PhD,Clayton.1992.PRC}.
However, the authors of~\cite{Edington.1966.NP,Clayton.1991.PhD} observed a difference in experimental data from the typical exponential shape of the bremsstrahlung spectrum previously measured by Edington and Rose in \cite{Edington.1966.NP}.
This point was under active discussion and could be informed by a possible explanation supported by calculations.
We emphasize that our new approach is applicable for analysis (i.e., this approach should be able to extract information about radius parameter $r_{R}$ of the proton-nucleus potential) of even conflicting experimental data.
So, we chose two experimental data sets \cite{Edington.1966.NP} and \cite{Clayton.1991.PhD,Clayton.1992.PRC} at the corresponding incident proton energies of 140~MeV and 145~MeV, and for the chosen angle between directions of the emitted photons and the incident protons which equals $90^{\circ}$.

\begin{figure}[htbp]
\centerline{\includegraphics[width=97mm]{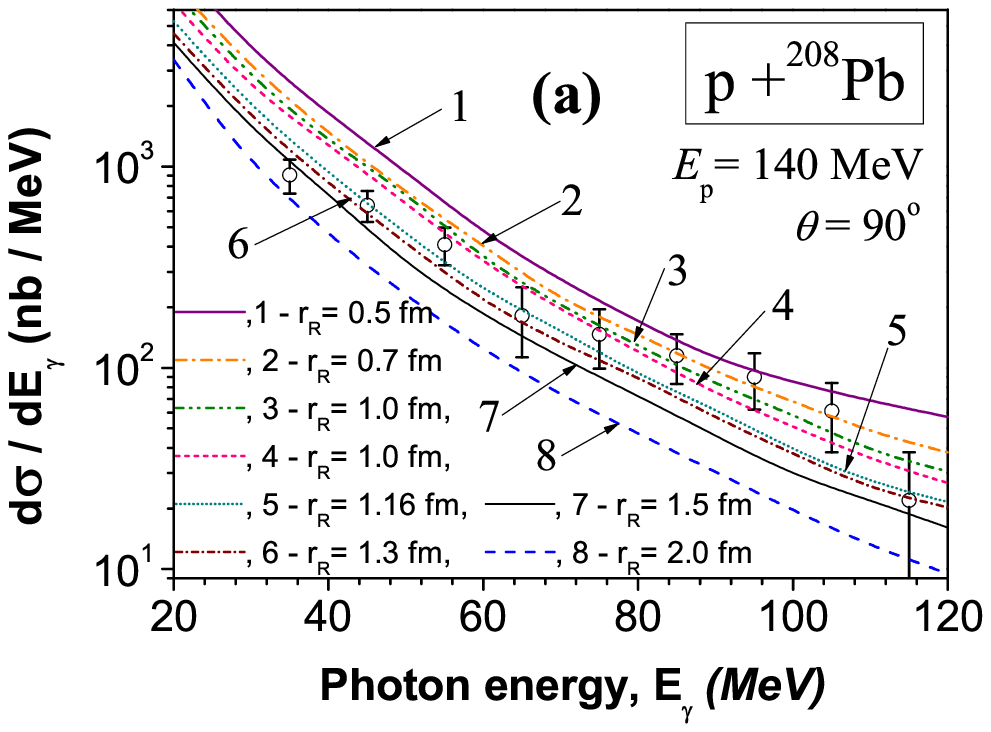}
\hspace{-9.5mm}\includegraphics[width=100mm]{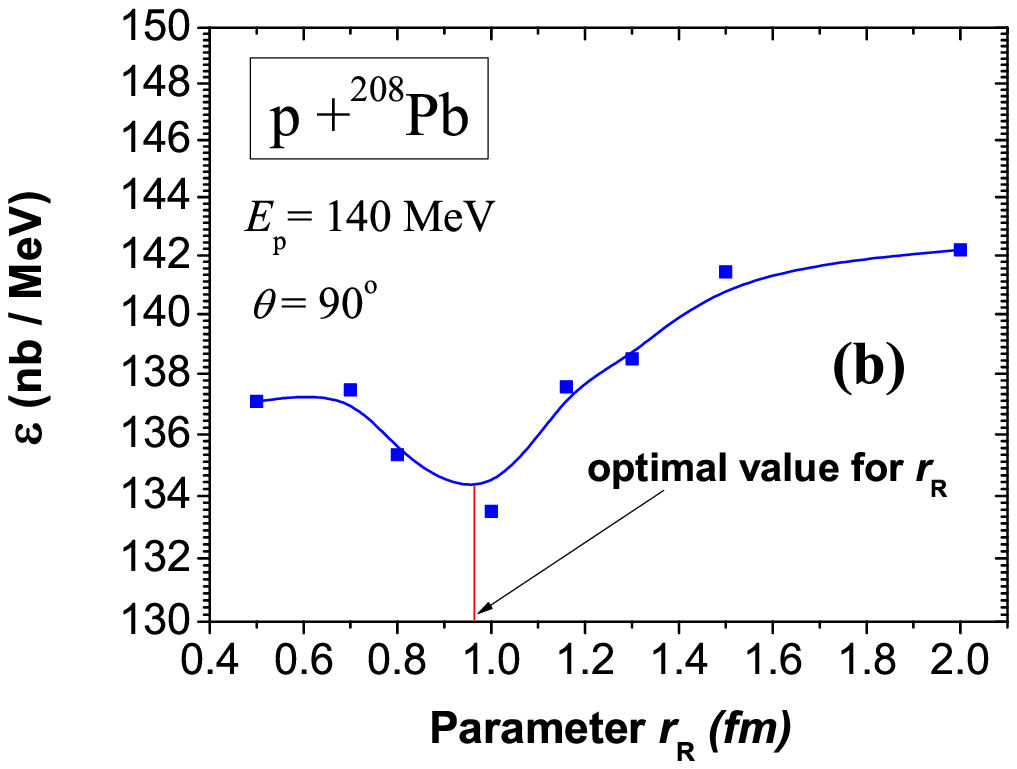}}
\vspace{-6mm}
\caption{\small (Color online)
(a) The bremsstrahlung cross sections for $p + ^{208}{\rm Pb}$ calculated on the basis of the first matrix element $p_{1}$ in Eq.~(\ref{eq.2.8.4}) at the incident proton energy of 140~MeV depending on the parameter $r_{R}$ in comparison with the experimental data of Edington and Rose~\cite{Edington.1966.NP} at $\theta=90^{\circ}$. One can see slow sensitivity of the calculated spectra from the parameter $r_{R}$. In particular, such a sensitivity should be present after normalization of all calculated spectra on the same experimental point.
(b) Estimated errors obtained by the method of minimization given by formula~(\ref{eq.2.11.3}) depending on values of the parameter $r_{R}$. One can see that there is a visible minimum which indicates the presence of optimal values for $r_{R}$, at which agreement between theory and experimental data is the highest.
\label{fig.1}}
\end{figure}

We shall clarify if the calculated spectrum is changed depending on the variation of the parameter $r_{R}$ (we use the approximation of $r_{C}=r_{R}$). We start our analysis from calculations of cross sections on the basis of the first matrix element $p_{1}$ given in Eq.~(\ref{eq.2.8.4}). Results of such calculations at 140~MeV of the proton energy
in comparison with experimental data~\cite{Edington.1966.NP} are presented in Fig.~\ref{fig.1}(a).
One can see that the spectra are slowly decreased with a decrease of this parameter. In order to obtain the accurate parameter, we compare the calculated spectrum with experimental data and calculate the error by formula~(\ref{eq.2.11.3}). We normalize each calculated curve on the same experimental point (we chose experimental data of $643\; {\rm nb} / ({\rm sr}\, {\rm MeV})$ at energy 45~MeV and an angle of $90^{\circ}$ taken from table~8 in \cite{Edington.1966.NP}, see p.~544).
In next Fig.~\ref{fig.1}(b) one can see the presence of a visible minimum in the tendency of the errors. This clearly confirms the stability in obtaining the minimal value for error by this method, and we find parameter $r_{R}$, for which an agreement between theory and experiment should be the most appropriate.

\begin{figure}[htbp]
\centerline{\includegraphics[width=99mm]{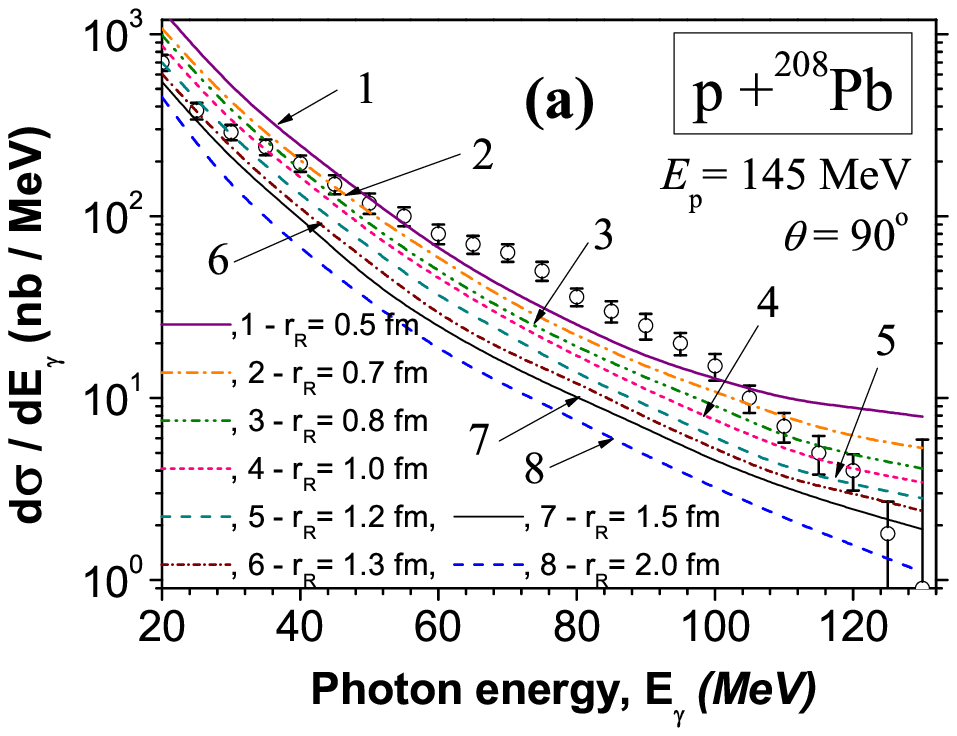}
\hspace{-9.5mm}\includegraphics[width=97.5mm]{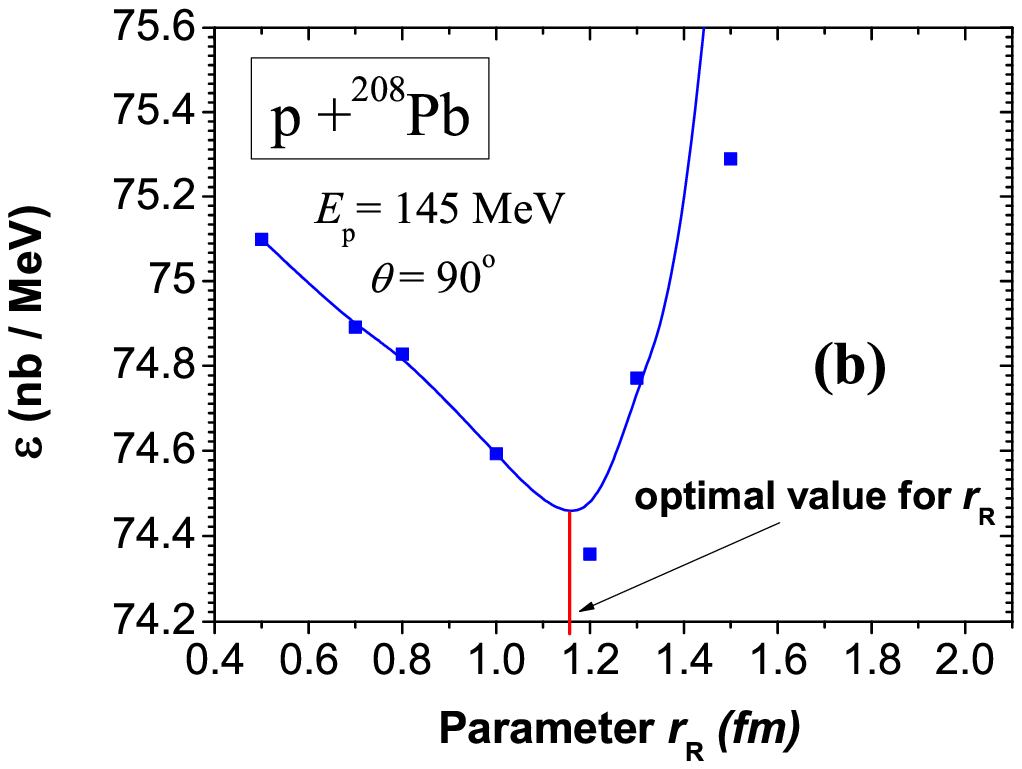}}
\vspace{-6mm}
\caption{\small (Color online)
(a) The bremsstrahlung cross sections for $p + ^{208}{\rm Pb}$ calculated on the basis of the first matrix element $p_{1}$ in Eq.~(\ref{eq.2.8.4}) at the incident proton energy of 145~MeV depending on the parameter $r_{R}$ in comparison with experimental data of Clayton \textit{et al.}~\cite{Clayton.1991.PhD,Clayton.1992.PRC} at $\theta=90^{\circ}$. Once again we obtain a slow sensitivity of the spectra on the parameter $r_{R}$.
One can see a clear difference between the logarithmic form of the calculated spectra and hump-shaped behavior in experimental data.
(b) Estimated errors obtained by the method of minimization depending on values of the parameter $r_{R}$. One can see that there is a visible minimum which indicates the presence of optimal values for $r_{R}$, at which agreement between theory and experimental data is the highest.
\label{fig.2}}
\end{figure}

Results for analysis of the bremsstrahlung in $p + ^{208}{\rm Pb}$ at the incident proton energy 145~MeV compared with experimental data of~\cite{Clayton.1991.PhD,Clayton.1992.PRC} are presented in Fig.~\ref{fig.2}. We observe the sensitivity of the calculated spectra on the parameter $r_{R}$ [see Fig.~\ref{fig.2}(a)], and find the clear minimum in dependence of function $\varepsilon$
given by Eq.~(\ref{eq.2.11.3}) on $r_{R}$ [see Fig.~\ref{fig.2}(b)].
After a comparison of the calculations with experimental data, we observe a difference between their shapes.
In particular, in the calculated spectra we obtain the shape of logarithmic type, which is typical for the bremsstrahlung emission in the $\alpha$ decay and fission (and here we obtained the most accurate agreement between calculations and the existed experimental data, see \cite{Maydanyuk.2006.EPJA,Maydanyuk.2008.EPJA,Giardina.2008.MPLA} and \cite{Maydanyuk.2010.PRC} for details).
At the same time, experimental data~\cite{Edington.1966.NP,Clayton.1991.PhD,Clayton.1992.PRC} have some different behavior with a little visible hump-shaped form inside 60--100 MeV [this is more clearly visible in Fig.~\ref{fig.2}(a)]. From here, we conclude that even experimental data for the proton-nucleus scattering and for the $\alpha$ decay and fission have a different behavior, that should be connected with some unclear physical reasons.

\begin{figure}[htbp]
\centerline{\includegraphics[width=97mm]{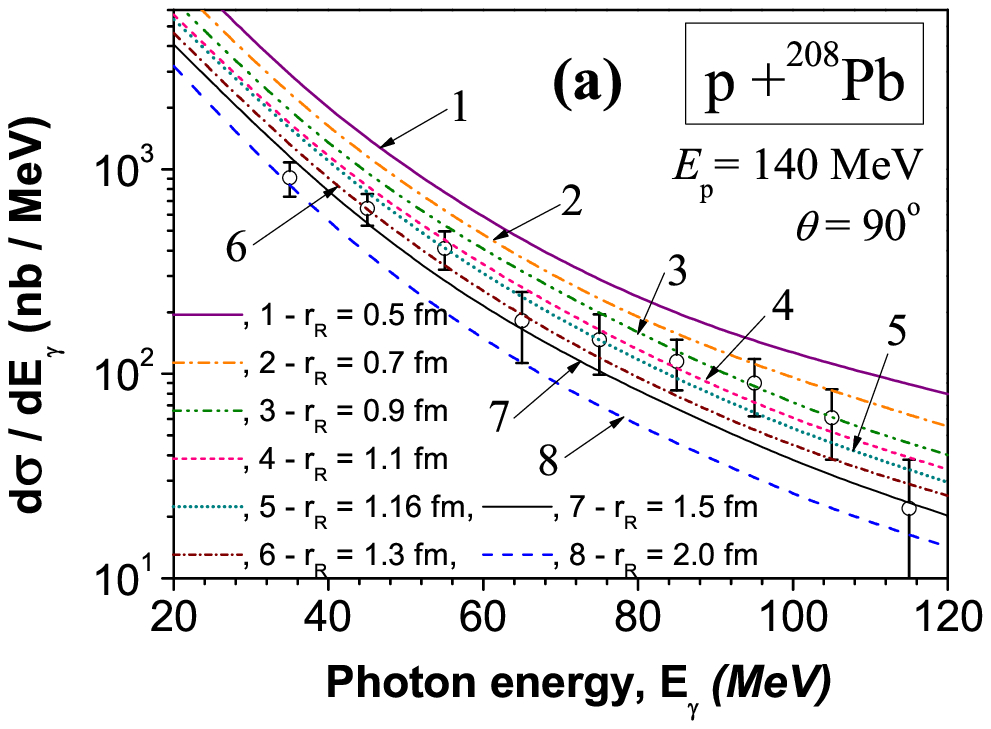}
\hspace{-9.5mm}\includegraphics[width=97mm]{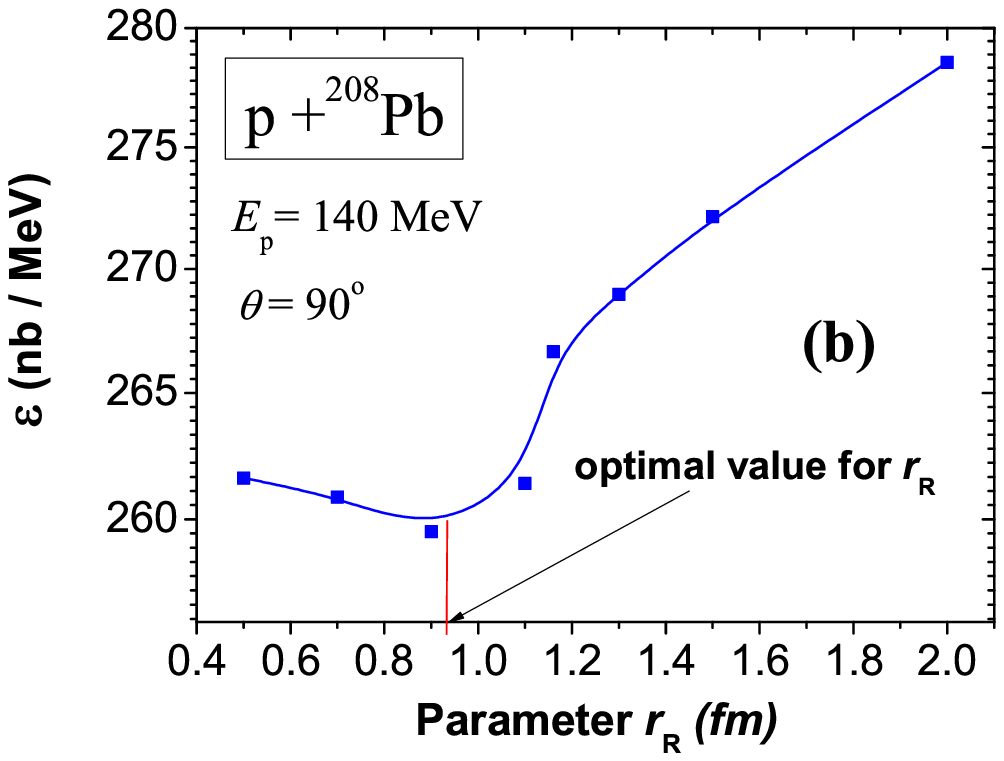}}
\vspace{-6mm}
\caption{\small (Color online)
(a) The calculated bremsstrahlung cross-sections for $p + ^{208}{\rm Pb}$ at 140~MeV on the basis of the second matrix element $p_{2}$ in Eq.~(\ref{eq.2.8.5}) at different values of the parameter $r_{R}$ in comparison with experimental data of Edington and Rose~\cite{Edington.1966.NP} at $\theta=90^{\circ}$.
(b) The estimated errors obtained by the method of minimization depending on values of the parameter $r_{R}$.
\label{fig.3}}
\end{figure}

\begin{figure}[htbp]
\centerline{\includegraphics[width=97mm]{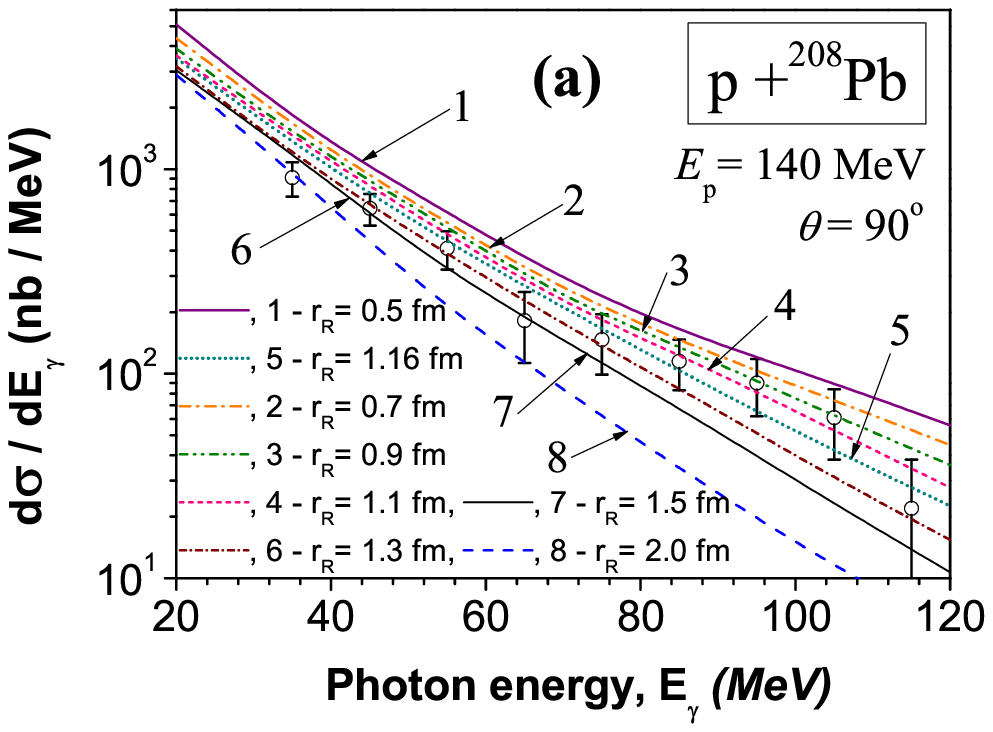}
\hspace{-9.5mm}\includegraphics[width=98mm]{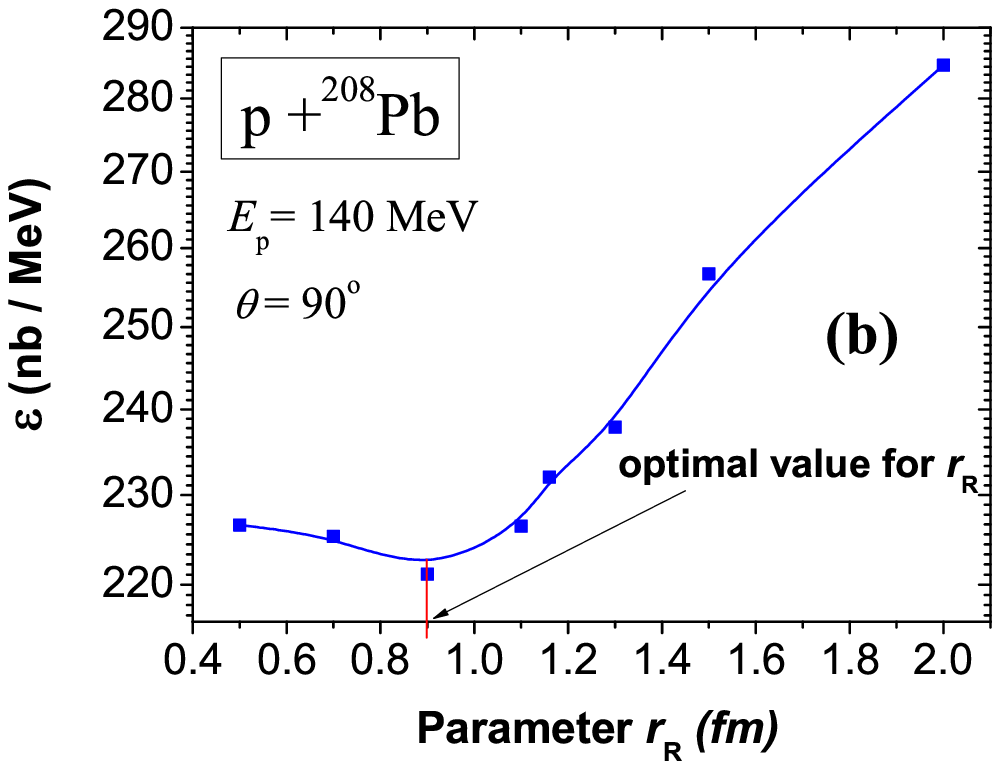}}
\vspace{-6mm}
\caption{\small (Color online)
(a) The calculated bremsstrahlung cross sections for $p + ^{208}{\rm Pb}$ at 140~MeV on the basis of the third matrix element $p_{3}$ in Eq.~(\ref{eq.2.8.6}) at different values of the parameter $r_{R}$ in comparison with experimental data of Edington and Rose~\cite{Edington.1966.NP} at $\theta=90^{\circ}$.
(b) The estimated errors obtained by the method of minimization depending on values of the parameter $r_{R}$.
\label{fig.4}}
\end{figure}

Our calculations of the bremsstrahlung spectra for $p + ^{208}{\rm Pb}$ at 140~MeV  on the basis of
the second matrix element $p_{2}^{\rm (dip)}$ in Eq.~(\ref{eq.2.8.5}) and the third matrix element $p_{3}$ in
Eq.~(\ref{eq.2.8.6}) give similar results (see Figs.~\ref{fig.3} and \ref{fig.4}).
We obtain the similar shapes of the logarithmic type for all calculated spectra.
The same results were obtained in calculations of the spectra at an energy of 145~MeV.
It is clear that all such calculated curves do not describe (and do not explain) the slowly visible hump-shaped plateau in the experimental data~\cite{Edington.1966.NP,Clayton.1991.PhD}
(see Fig.~\ref{fig.2}).
Errors for all new calculations are a little larger than for the calculations on the basis of the first matrix element $p_{1}$.
However, the difference between the parameters $r_{R}$ for each case obtained by the minimization is sufficiently small
(see Table~\ref{table.2}).
\begin{table}
\begin{tabular}{|c|c|c|c|c|} \hline
  Matrix element $p_{i}$ used in & \multicolumn{2}{|c|}{Analysis of data~\cite{Edington.1966.NP}} &
    \multicolumn{2}{|c|}{Analysis of data~\cite{Clayton.1991.PhD,Clayton.1992.PRC}} \\
 \cline{2-5}

  calculations of cross sections &
  parameter $p_{R}$, fm & error $\varepsilon$ &
  parameter $p_{R}$, fm & error $\varepsilon$ \\ \hline
  Matrix element $p_{1}$ &
  0.97 & 134 & 1.15 & 74.5 \\
  Matrix element $p_{2}^{\rm (dip)}$ &
  0.93 & 261 & 1.16 & 95.5 \\
  Matrix element $p_{3}$ &
  0.90 & 223 & 1.17 & 83.5 \\
\hline
\end{tabular}
\caption{The values for parameter $p_{R}$ obtained by minimization and the corresponding errors.
\label{table.2}}
\end{table}

\begin{figure}[htbp]
\centerline{\includegraphics[width=97mm]{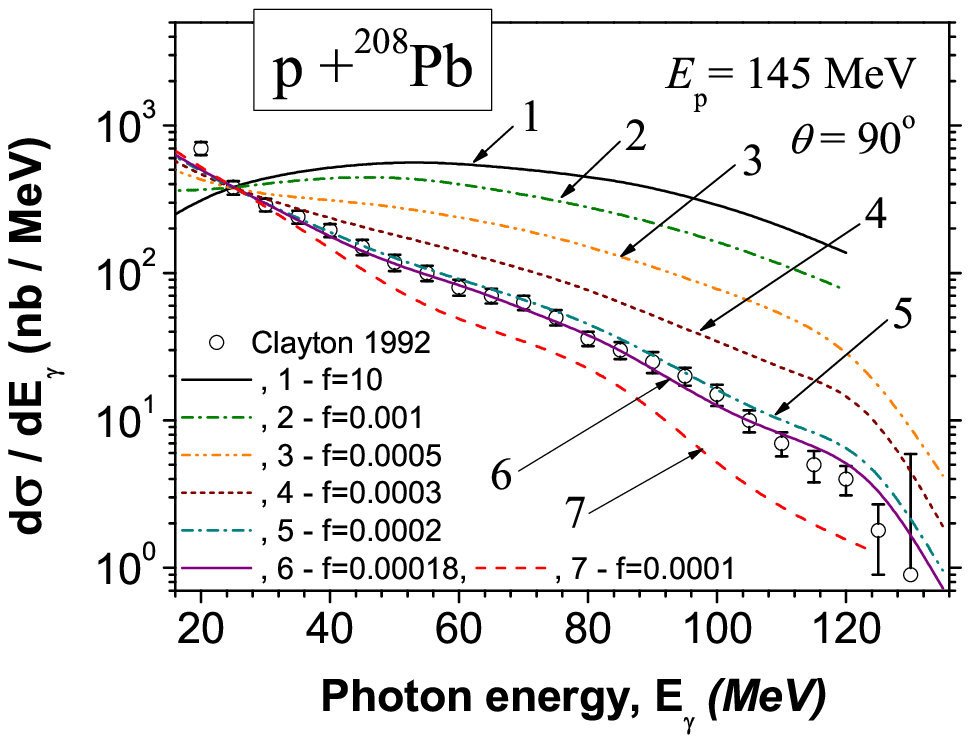}}
\vspace{-2mm}
\caption{\small (Color online)
The bremsstrahlung cross sections for $p + ^{208}{\rm Pb}$ at 145~MeV of the proton energy calculated on the basis of the last matrix element $p_{4}$ in Eq.~(\ref{eq.2.9.9}) depending on the parameter $f$ in comparison with the experimental data of Clayton \textit{et al.}~\cite{Clayton.1992.PRC} at $\theta=90^{\circ}$.
The best agreement between theory and experimental data is obtained for the factor $f=0.00018$
(used parameters of potential: $r_{R} = r_{C} = 1.17$~fm).
\label{fig.5}}
\end{figure}

The inclusion of the last matrix element $p_{4}$ from Eq.~(\ref{eq.2.9.8}) to calculations changes the bremsstrahlung spectrum. In particular, the main part of the spectrum is transformed from the logarithmic type on the hump-shape.
Such an interesting picture is observed more clearly in the experimental data~\cite{Clayton.1991.PhD,Clayton.1992.PRC}.
However, a general tendency of the calculated spectrum and its slope are sensitive to the relative contribution of the term $p_{3}$ in Eq.~(\ref{eq.2.8.6}) into a full matrix element $p_{4}$ [which we determine via factor $f$, see Eq.~(\ref{eq.2.9.10})].
Such a contribution is determined by weight amplitudes of the proton-nucleus wave function in different states with different quantum numbers and their interference (see \cite{Maydanyuk.2012.PRC} for details and formalism).
Thus, it could be interesting to analyze how the spectrum is changed depending on such a contribution.
Results of such calculations in comparison with experimental data \cite{Clayton.1992.PRC}
are presented in Fig.~\ref{fig.5}.
From here we find a nice agreement between our calculations and experimental data at $f=0.00018$ which is observed practically inside the whole energy region of the emitted photons where experimental data are available (with a minor exception of the first data point and third to last one). This improves our previous results describing the emitted photons in the $p + ^{208}{\rm Pb}$ reaction given in~\cite{Maydanyuk.2012.PRC} after the inclusion of the consideration of the nucleus as a system of many nucleons and their dynamic properties.

From here we conclude that the role of the dynamics of nucleons inside the nucleus and their connection with spin properties of the incident proton is essential (for lower and middle energy of the emitted photons).
Now it becomes clear that in the study of the bremsstrahlung emission in the $\alpha$ decay we did not have such humped-shaped spectra, because the $\alpha$ particle has zero spin and, so, the corresponding contribution is zero also. So, the difference between the experimental spectra for the proton-nucleus scattering and the $\alpha$ decay is explained on the physical basis.

Detailed consideration of the resulting spectrum shows the presence of a slowly decreasing hump-shaped plateau from 40 up to 120 MeV of energies of the emitted photons.
However, the lowest energy region in the spectrum (below 40~MeV) has the logarithmic shape. Such a separation can be explained by photons at lower energies being emitted in results of coherent processes (i.e., such photons are formed as a result of interactions between the scattering proton and nucleus as a whole object, without internal consideration of the many-nucleon structure of the nucleus). For higher photon energies, we obtain the hump-shaped spectrum which can be explained by the essential role in the emission of interactions between the scattering proton and momenta of nucleons of the nucleus (i.e., noncoherent processes).

\begin{figure}[htbp]
\centerline{\includegraphics[width=97mm]{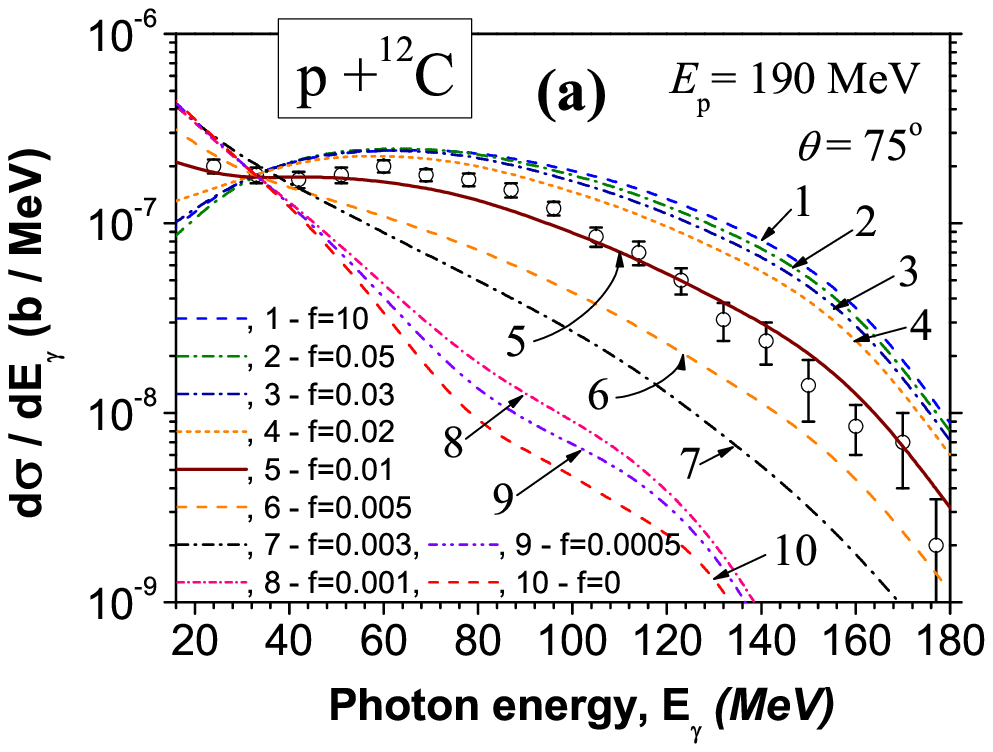}
\hspace{-9.5mm}\includegraphics[width=97mm]{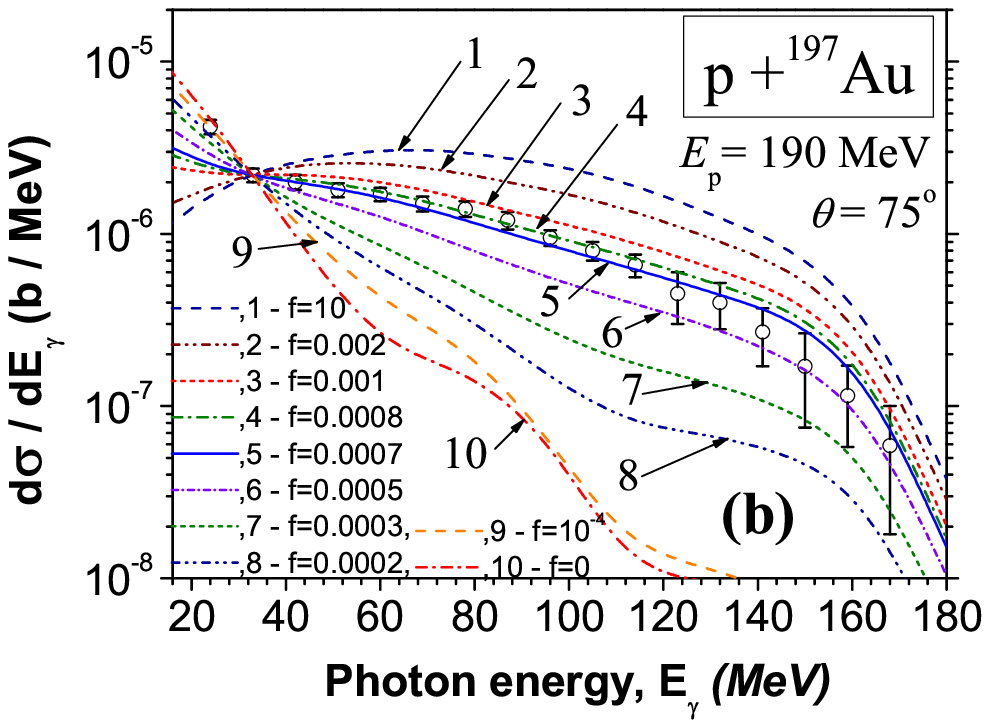}}
\vspace{-6mm}
\caption{\small (Color online)
The bremsstrahlung cross-sections for $p + ^{12}{\rm C}$ (a) and $p + ^{197}{\rm Au}$ (b) at 190~MeV of the proton energy calculated on the basis of the last matrix element $p_{4}$ in Eq.~(\ref{eq.2.9.9}) depending on the parameter $f$
in comparison with
the experimental data of van~Goethem \textit{et al.} \cite{Goethem.2002.PRL}
at the angle of $\theta=75^{\circ}$
(all calculated spectra are normalized on the second point of experimental data,
used parameters of potential are $r_{R} = r_{C} = 0.95$~fm).
Once again, we obtain a visible sensitivity of the spectra on the parameter $f$ characterized contribution of the
emission caused by the dynamics of the nucleons of the nucleus.
The best agreement between theory and experimental data is observed for
$p + ^{12}{\rm C}$ at the factor $f=0.01$ [see purple solid line 5 in the figure (a)]
and $p + ^{197}{\rm Au}$ at the factor $f=0.0007$ [see blue solid line 5 in the figure (b)].
This result demonstrates that the inclusion of the dynamics of nucleons and its connection with spin properties of the scattering proton into the model allows us to describe (and, so, to explain)
the existing plateau in the spectra of the bremsstrahlung photons.
\label{fig.6}}
\end{figure}
\begin{figure}[htbp]
\centerline{\includegraphics[width=97mm]{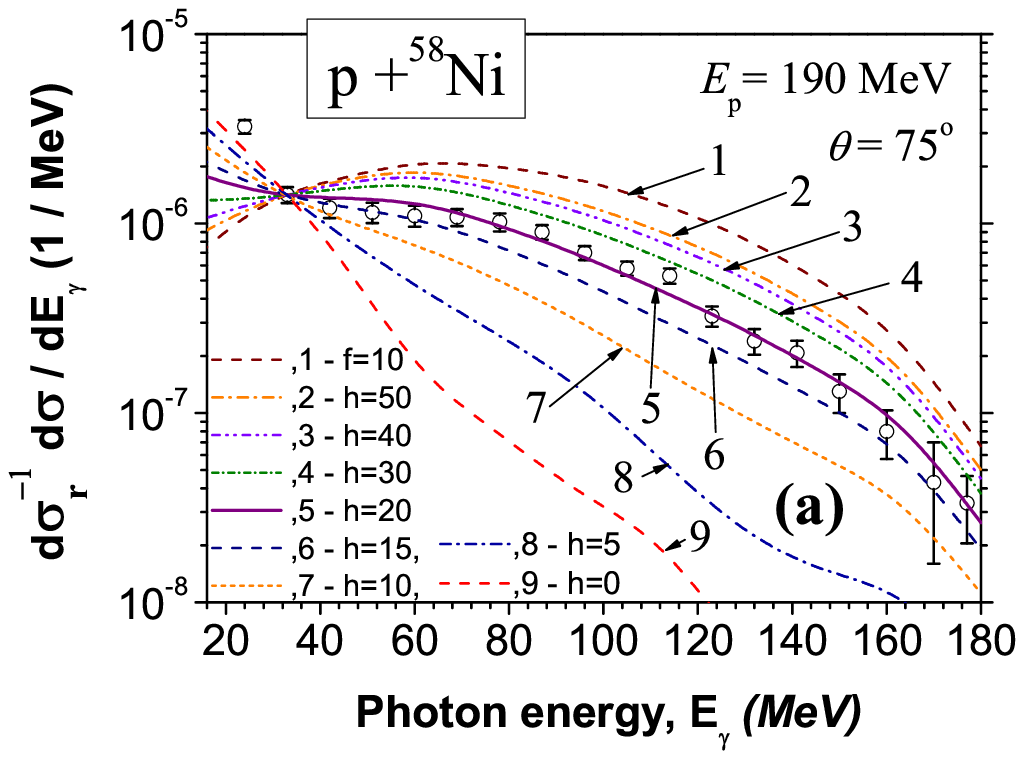}
\hspace{-9.5mm}\includegraphics[width=97mm]{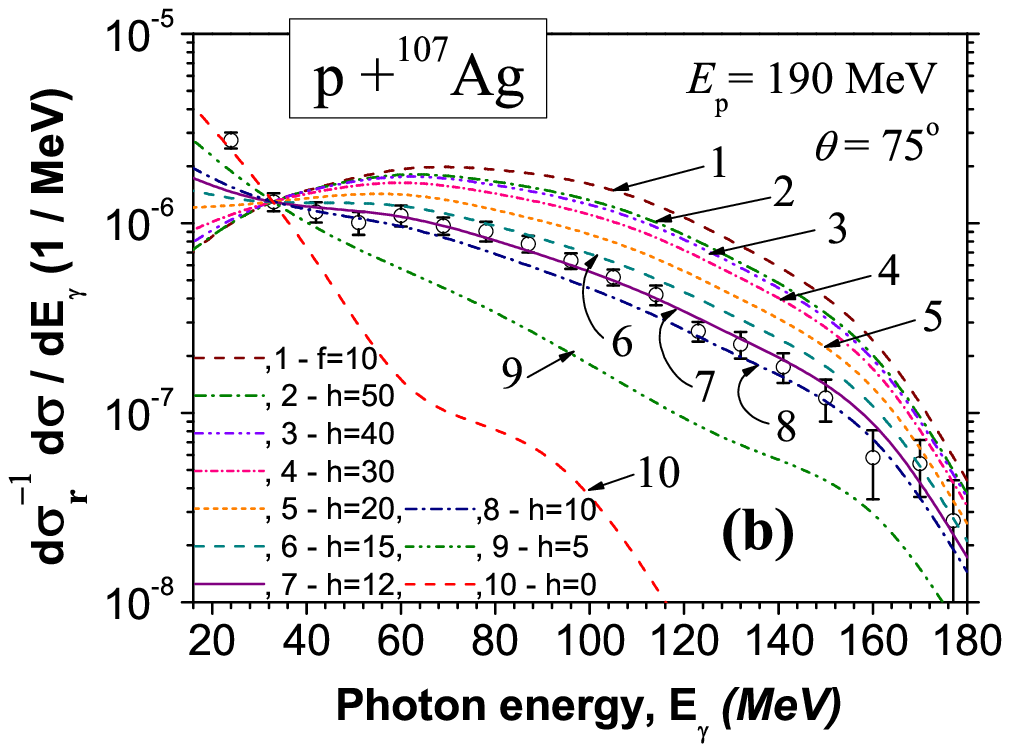}}
\vspace{-6mm}
\caption{\small (Color online)
The bremsstrahlung cross sections for $p + ^{58}{\rm Ni}$ (a) and $p + ^{107}{\rm Ag}$ (b)
at 190~MeV of the proton energy calculated on the basis of the last matrix element $p_{4}$ in Eq.~(\ref{eq.2.9.9}) depending on the parameter $f$
in comparison with
the experimental data of van~Goethem \textit{et al.}~\cite{Goethem.2002.PRL}
at the angle of $\theta=75^{\circ}$
(the experimental data are normalized on the geometrical cross section,
all calculated spectra are normalized on the second point of experimental data,
used parameters of potential are $r_{R} = r_{C} = 0.95$~fm,
in all figures we use the renormalized factor $h = f \times 10^{4}$).
The best agreement between theory and experimental data is observed for
$p + ^{58}{\rm Ni}$ at the factor $f=0.002$ [see purple solid line 5 in the figure (a)]
and $p + ^{107}{\rm Ag}$ at the factor $f=0.0012$ [see purple solid line 7 in the figure (b)].
\label{fig.7}}
\end{figure}

The careful measurements of the bremsstrahlung emission in the proton nucleus scattering were done by the TAPS collaboration \cite{Goethem.2002.PRL} and we must include them in our analysis.
In Fig.~\ref{fig.6} we present the results of our calculations of the bremsstrahlung spectra
for $p + ^{12}{\rm C}$ (a) and $p + ^{197}{\rm Au}$ (b)
at $E_{\rm p}=190$~MeV in comparison with these experimental data
(see Figs.~2 and 3 in~\cite{Goethem.2002.PRL}).
In Fig.~\ref{fig.7} we add our calculations of the bremsstrahlung spectra
for $p + ^{58}{\rm Ni}$ (a) and $p + ^{107}{\rm Ag}$ (b)
at $E_{\rm p}=190$~MeV in comparison with experimental data~\cite{Goethem.2002.PRL} which are normalized on the geometrical cross section $\sigma_{r} = 1.44\, \pi A^{2/3}\; \mbox{\rm fm}^{2}$ with $A$ the target mass number
(see Fig.~1 in~\cite{Goethem.2002.PRL} for the data and text in that paper for details).
It can be seen that the inclusion of the contribution caused by the dynamics of nucleons of the nucleus into the model and calculations allows to describe these experimental data enough well inside practically the whole energy region of the emitted photons
(with the possible exception of the first data point for some reactions).

Agreement between our model and experimental data for all considered nuclei shows that the contribution of the emission of photons into the full spectrum, caused by a connection between the internal momenta of nucleons inside the nucleus and the spin properties of the scattering proton, is essential,
and its presence is confirmed (proven) by the experimental data.
On such a basis we can consider such an emission of the bremsstrahlung photons as some new type of emission, which can be named
\emph{the bremsstrahlung emission on the basis of spin-internal nucleons momenta interactions}.
Introduction of such a type of emission allows us for the first time to explain the origin (presence) of the hump-shaped plateau in the bremsstrahlung spectra for the proton-nucleus scattering, and, at the same time, the absence of such a plateau in the spectra for the $\alpha$ decay
(see data from~\cite{Boie.2007.PRL,Boie.2009.PhD} and
\cite{Maydanyuk.2006.EPJA,Maydanyuk.2008.EPJA,Giardina.2008.MPLA}).
Also we find that the optical model of the scattering of protons off nuclei does not include such a term of interactions. Therefore, our results on the study of the bremsstrahlung photons in the proton-nucleus scattering indicate the recommendation to generalize the optical model with the involution of such a spin-momenta term.

\begin{figure}[htbp]
\centerline{\includegraphics[width=97mm]{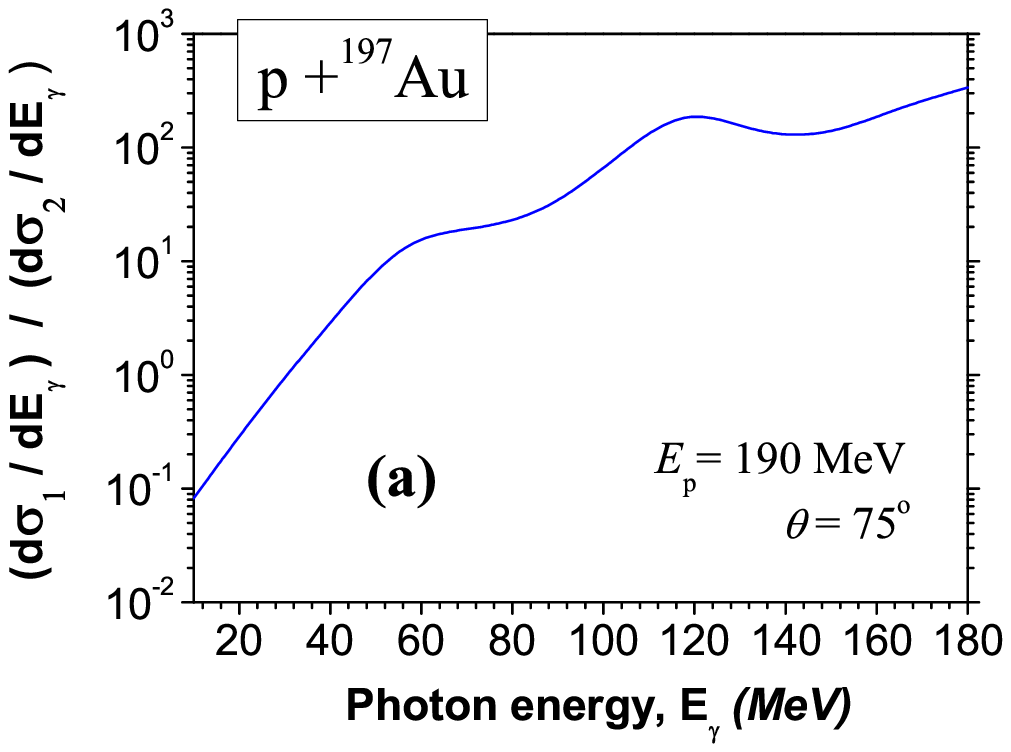}
\hspace{-9.1mm}\includegraphics[width=97mm]{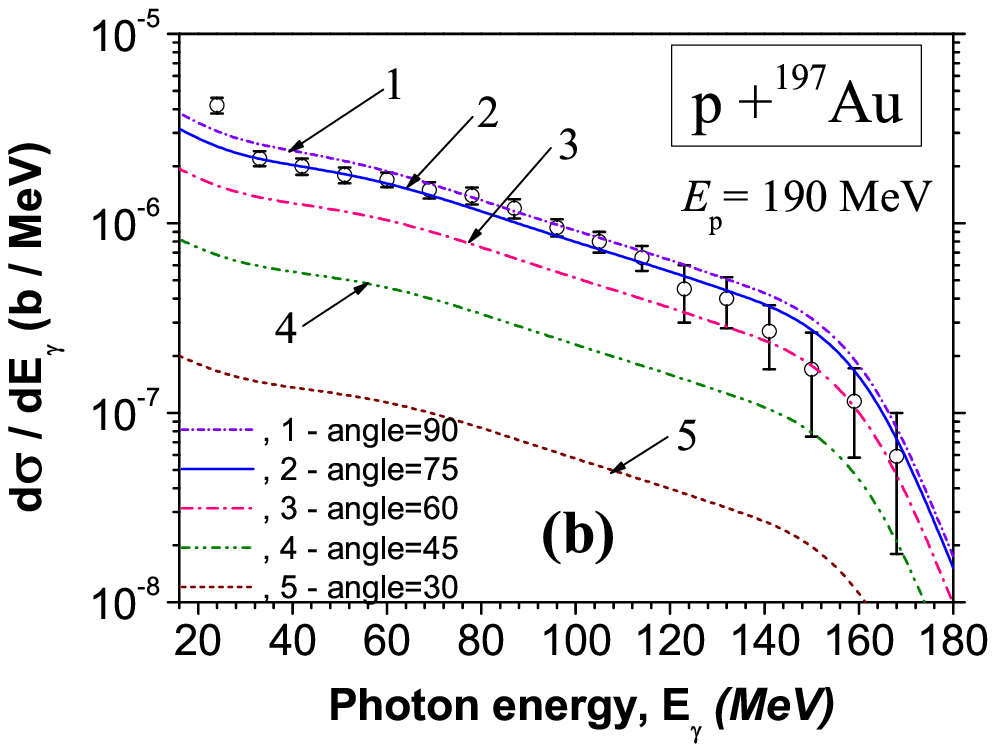}}
\vspace{-6mm}
\caption{\small (Color online)
(a) The ratio between the incoherent emission (formed by interactions between the scattering proton and internal momenta of nucleons of nucleus, denoted as $d \sigma_{1}/ dE_{\gamma}$) and the coherent emission (formed by interaction between the scattering proton and nucleus as whole without its internal many-nucleon structure, denoted as $d \sigma_{2}/ dE_{\gamma}$) depending on the energy of emitted photons
for $p + ^{197}{\rm Au}$ at $E_{\rm p}=190$~MeV and the angle of $\theta=75^{\circ}$
[used parameters of potential are $r_{R} = r_{C} = 0.95$~fm,
$f = 0.0007$ according to results in Fig.~\ref{fig.6}(b)].
(b) The angular distribution of the emitted photons for $p + ^{197}{\rm Au}$ at $E_{\rm p}=190$~MeV in comparison with the experimental data of van~Goethem \textit{et al.}~\cite{Goethem.2002.PRL}
measured at the angle of $\theta=75^{\circ}$
[$f = 0.0007$ according to results in Fig.~\ref{fig.6}(b)].
\label{fig.8}}
\end{figure}

One can estimate how much the incoherent emission (formed by interactions between the scattering proton and internal momenta of nucleons of the nucleus) is changed concerning coherent emission (formed by the interaction between the scattering proton and nucleus as a whole without the consideration of its internal many-nucleon structure). Such characteristic can be determined via the ratio between squares of the corresponding matrix elements,
i.e., as $|f \cdot p_{4}|^{2} /\, |p_{1}|^{2}$.
Such calculations for the reaction $p + ^{197}{\rm Au}$ are given in Fig.~\ref{fig.8}(a).
From such results one can see that in the photon energy region up to 30~MeV the coherent emission is more intensive [ratio is less than unity in Fig.~\ref{fig.8}(a)]. However, at higher photon energies the intensity of the incoherent emission is increased and becomes essential at high energies. A similar situation is present for the other studied reactions above.
In Fig.~\ref{fig.8}(b) we add our calculations for the angular distribution of the emitted photons for $p + ^{197}{\rm Au}$ at $E_{\rm p}=190$~MeV. As differences between the calculated spectra are larger than experimental errors for the angle of $\theta=75^{\circ}$ given by van~Goethem \textit{et al.}~\cite{Goethem.2002.PRL}, we suppose such results could be used by experimentalists as a possible future test of our model, the obtained spectra and analysis of peculiarities of the bremsstrahlung photons.


\section{Conclusions and perspectives
\label{sec.conclusions}}

In this paper we presented the new developments of our model of the bremsstrahlung emission
(presented previously
in~\cite{Maydanyuk.2011.JPG,Maydanyuk.2012.PRC,Maydanyuk.2003.PTP,Maydanyuk.2006.EPJA,Maydanyuk.2008.EPJA,%
Giardina.2008.MPLA,Maydanyuk.2009.TONPPJ,Maydanyuk.2009.JPS,Maydanyuk.2009.NPA,Maydanyuk.2010.PRC})
which accompanies the scattering of protons off nuclei at low and intermediate energies of the emitted photons.
In the analysis we studied the scattering of $p + ^{208}{\rm Pb}$ at the proton incident energies of $E_{\rm p}=140$ and 145~MeV,
the scattering of $p + ^{12}{\rm C}$, $p + ^{58}{\rm Ni}$, $p + ^{107}{\rm Ag}$, and $p + ^{197}{\rm Au}$
at the proton incident energy of $E_{\rm p}=190$~MeV.
We emphasized the extraction of new information about proton--nucleus interactions from the analysis of existing bremsstrahlung experimental data \cite{Edington.1966.NP,Clayton.1991.PhD,Clayton.1992.PRC,Goethem.2002.PRL}.
Note the following:
\begin{itemize}

\vspace{1.2mm}
\item
The calculated cross sections on the basis of the first, second and third matrix elements $p_{1}$, $p_{2}^{\rm (dip)}$ and $p_{3}$ given in Eqs.~(\ref{eq.2.8.4}), (\ref{eq.2.8.5}), and (\ref{eq.2.8.6}) have similar shapes of the logarithmic type.
Such calculations are found to be in some agreement with experimental data~\cite{Edington.1966.NP} for $p + ^{208}{\rm Pb}$ at $E_{\rm p}=140$~MeV.
However, they disagree with experimental data \cite{Clayton.1991.PhD,Clayton.1992.PRC} for $p + ^{208}{\rm Pb}$ at $E_{\rm p}=145$~MeV and \cite{Goethem.2002.PRL} for $p + ^{12}{\rm C}$, $p + ^{58}{\rm Ni}$, $p + ^{107}{\rm Ag}$ and $p + ^{197}{\rm Au}$ at $E_{\rm p}=190$~MeV.

\vspace{1.2mm}
\item
Inclusion of the last matrix element $p_{4}$ in Eq.~(\ref{eq.2.9.8}) into the calculations changes the bremsstrahlung spectrum.
The full spectrum has the hump-shaped plateau inside the middle energy region and then it decreases to the kinematic limit of the photons energies. The lowest energy region in the spectrum has a logarithmic shape.
Such a separation is explained by the fact that at lower energies photons are emitted from the results of coherent processes (i.e., interactions between the scattering proton and nucleus as a whole object, without internal consideration of many-nucleon structure of the nucleus),
while for higher photon energies the role of noncoherent processes (interactions between the scattering proton and momenta of nucleons of the nucleus) is essential in emission.
In such a way, we obtain a nice agreement inside practically the whole energy region of the emitted photons (with the possible exception of the first data point) with experimental data~\cite{Clayton.1991.PhD,Clayton.1992.PRC} for $p + ^{208}{\rm Pb}$ at $E_{\rm p}=145$~MeV and
\cite{Goethem.2002.PRL} for $p + ^{12}{\rm C}$, $p + ^{58}{\rm Ni}$, $p + ^{107}{\rm Ag}$ and $p + ^{197}{\rm Au}$ at $E_{\rm p}=190$~MeV.
This result improves our previous calculations in the description of the emitted photons in the $p + ^{208}{\rm Pb}$ reaction given in~\cite{Maydanyuk.2012.PRC} after the inclusion of the nucleus as a system of many nucleons and their dynamical properties taken into consideration and included in the model.
From here we conclude that the role of the dynamics of nucleons inside the nucleus and its connection with spin properties of the incident proton is really essential.
This shows a perspective to further study the dynamics of nucleons inside the nucleus experimentally.

\vspace{1.2mm}
\item
By using this approach we explain the difference between the bremsstrahlung spectra in the proton-nucleus scattering and the bremsstrahlung spectra in the $\alpha$ decay and heavy-ion reactions.
Note that in calculations of the bremsstrahlung emission in the $\alpha$ decay we obtained the shape of the logarithmic type without such a humped-shaped form
(here we achieved the most accurate agreement between our calculations~\cite{Maydanyuk.2006.EPJA,Maydanyuk.2008.EPJA,Giardina.2008.MPLA} and the existing experimental data~\cite{Boie.2007.PRL,Boie.2009.PhD,Maydanyuk.2008.EPJA,Giardina.2008.MPLA}, a similar situation is observed also for fission~\cite{Maydanyuk.2012.PRC}).
Now it becomes clear that the absence of such humped-shaped spectra for the $\alpha$ decay is explained by
zero spin of the $\alpha$ particle, and the resulting zero matrix elements in Eqs.~(\ref{eq.2.7.5}) and (\ref{eq.2.8.5}).
So, a difference between tendencies of the experimental bremsstrahlung spectra for the proton-nucleus scattering~\cite{Clayton.1991.PhD,Clayton.1992.PRC} and the $\alpha$ decay~\cite{Boie.2007.PRL,Boie.2009.PhD,Maydanyuk.2008.EPJA,Giardina.2008.MPLA} is explained on the physical basis.
However, at higher energies of photons (in comparison with energies for the $\alpha$-decay), in interactions between the $\alpha$ particles and nuclei (as in the $\alpha$-particle--nucleus scattering) the internal structure of the $\alpha$ particle seems to play more important role, that would produce some nonminor incoherent contribution to the full bremsstrahlung spectrum.

\vspace{1.2mm}
\item
We observe a slow dependence of the calculated bremsstrahlung spectra on the radius-parameter $r_{R}$ used in the definition of the proton-nucleus potential in Eqs.~(\ref{eq.2.11.1})--(\ref{eq.2.11.2}) (we use $r_{C} = r_{R}$).
Analysis shows the following:
\begin{itemize}
\item
The spectra are slowly decreased with a decrease of this parameter.

\item
In order to get the proper radius parameter, we compare each calculated spectrum with experimental data and calculate the function of errors $\varepsilon$ by formula~(\ref{eq.2.11.3}). We normalize each calculated curve on the same experimental point.
The resulting dependence of the function of errors on the radius parameter has a clear visible minimum [for example, see Figs.~\ref{fig.1}(b), \ref{fig.2}(b)].
This confirms the stability in obtaining the minimal value for the function of errors.
We find $r_{R}=0.90$~fm in the analysis of~\cite{Edington.1966.NP} at $E_{\rm p}=140$~MeV and $r_{R}=1.17$~fm in the analysis of~\cite{Clayton.1991.PhD,Clayton.1992.PRC} at $E_{\rm p}=145$~MeV.
Taking into account the best agreement between calculations and experimental data~\cite{Clayton.1991.PhD,Clayton.1992.PRC} (see Fig.~\ref{fig.5}) after the inclusion of the matrix element $p_{4}$, we choose the case of $r_{R}=1.17$~fm
(this result is in agreement with results~\cite{Becchetti.1969.PR} obtained from the fitting procedure in scattering, which is $r_{R} = 1.17\;{\rm fm}$ and $r_{c} = 1.22\;{\rm fm}$).
Analyzing the experimental data~\cite{Goethem.2002.PRL} at $E_{\rm p}=190$~MeV, we obtain
$r_{R} = 1.09\;{\rm fm}$ for $p + ^{12}{\rm C}$,
$r_{R} = 1.11\;{\rm fm}$ for $p + ^{58}{\rm Ni}$,
$r_{R} = 1.12\;{\rm fm}$ for $p + ^{107}{\rm Ag}$ and
$r_{R} = 1.08\;{\rm fm}$ for $p + ^{197}{\rm Au}$.
\end{itemize}

\vspace{1.2mm}
\item
We have estimated how much the incoherent emission (formed by interactions between the scattering proton and internal momenta of nucleons of the nucleus) is changed by coherent emission (formed by an interaction between the scattering proton and nucleus as a whole without consideration of its internal
many-nucleon structure).
According to an analysis for the reaction $p + ^{197}{\rm Au}$ at $E_{\rm p}=190$~MeV and the angle of $\theta=75^{\circ}$,
in the photon energy region up to 30~MeV the coherent emission is more intensive [ratio is less than unity in Fig.~\ref{fig.8}(a)].
However, at higher photon energies the intensity of the incoherent emission is increased [ratio is larger than unity in Fig.~\ref{fig.8}(a)] and becomes essential at high energies [ratio is close to 336 in Fig.~\ref{fig.8}(a)].
A similar situation is present for the other studied reactions in this paper.
From such results it follows that the role of the internal dynamics of nucleons in the nucleus is essential in the high-energy bremsstrahlung emission in the scattering of the particles with nonzero spin, and it only increases with increasing photon energy.
Also we see that our approach in the inclusion of the incoherent component of emission has some similar logic as the method provided by Nakayama and Bertsch (for example, see Ref.~\cite{Nakayama.1989.PRCv40}).

\vspace{1.2mm}
\item
We add our predictions for the angular distribution of the emitted photons for $p + ^{197}{\rm Au}$ at $E_{\rm p}=190$~MeV [see Fig.~\ref{fig.8}(b)]. As differences between the spectra calculated at the different angles are larger than experimental errors for the angle of $\theta=75^{\circ}$ given by van~Goethem \textit{et al.}~\cite{Goethem.2002.PRL}, we suppose such results could be used by experimentalists to test our model, the obtained spectra and analysis of peculiarities of the bremsstrahlung photons. Such new experiments will allow us to confirm our information about the role of the internal dynamics of nucleons in the full bremsstrahlung emission. In particular, we propose such possible angular experimental measurements could be performed on the PROTEUS C-235 proton cyclotron (which produces a proton beam with energies 70--230~MeV) at the Henryk Niewodniczanski Institute of Nuclear Physics in Krakow with the use of the HECTOR array for the angular registering of the bremsstrahlung photons (see Ref.~\cite{Maj.1994.NPA} for details, also see research \cite{Maj.1992.PL,Bracco.1995.PRL,Camera.1999.PRC,Tveter.1996.PRL,
Benzoni.1996.PLB,Maj.2004.NPA,Kmiecik.2004.PRC,Wieland.2009.PRL} where such a facility was used).

\vspace{1.2mm}
\item
Results presented above answer the question in~\cite{Kopitin.1997.YF} as to whether there is a sense to put forces and develop potential models (taking into account many nucleons and collective effects) for a description of bremsstrahlung in proton-nucleus scattering and nucleus-nucleus collisions. The new method, with its improvements in accuracy and stability, provides an effective tool to investigate the new detailed information about proton-nucleus interactions and mechanisms of photon emission.

\vspace{1.2mm}
\item
After achieving an agreement between experimental data and calculations, we can now observe (for the first time) the presence of very tiny oscillations in the bremsstrahlung experimental data. One can suppose that it can be connected with some less visible physical effects or peculiarities of the proton-nucleus scattering process. Here, we can recall the hypothesis presented by Eremin, Olkhovsky and Giardina
(see some related papers \cite{Olkhovsky.1968.NCA,Olkhovsky.1969.NCA,Olkhovsky.1970.LNC,Olkhovsky.1974.NCA,
Olkhovsky.1992.PR,Olkhovsky.2004.PR,Olkhovsky.2010.PEPAN,Maydanyuk.2003.PhD})
many years before about the possible connection of such oscillations in the bremsstrahlung spectra and tunneling time through the barrier.
Adding our research in the $\alpha$ decay problem, we can now conclude that such a phenomenon is general enough for the nuclear reactions.
\end{itemize}

\section*{Acknowledgments
\label{sec.acknowledgments}}

S.~P.~M. is grateful to Dr.~Andrii~I.~Steshenko for his insight and support in understanding the microscopic approaches developments.
The authors appreciate the support provided by Prof.~Adam Maj and Dr.~Maria Kmiecik in understanding the experimental facilities at measurements of the bremsstrahlung photons in nuclear reactions.
S.~P.~M. thanks Institute of Modern Physics of Chinese Academy of Sciences for its warm hospitality and support.
This work was supported by the Major State Basic Research Development Program in China (No. 2015CB856903), the National Natural Science Foundation of
China (Grant Nos. 11035006 and 11175215), and
the Chinese Academy of Sciences fellowships for researchers from developing countries (No. 2014FFJA0003).


\appendix
\section{The form factor of the nucleus
\label{sec.app.1}}

\subsection{Form factor of the system composed from many nucleons
\label{sec.app.1.1}}

Let us consider the electromagnetic form factor of the nuclear system composed from nucleons with number $A$ (in this appendix we shall omit the bottom index $A$ for all variables,
indicating belonging of nucleons to the nucleus):
\begin{equation}
  Z_{A} (\mathbf{k}) =
  \Bigl\langle\: \psi_{\rm nucl, f} (1 \ldots A )\: \Bigl|\,
    \displaystyle\sum\limits_{s=1}^{A} Z_{s}\, \displaystyle\frac{m_{\rm p}}{m_{s}}\;
    e^{-i \mathbf{k} \cdot \rhobfsm_{s} }\,
  \Bigr|\: \psi_{\rm nucl, i} (1 \ldots A )\: \Bigr\rangle.
\label{eq.app.1.1.1}
\end{equation}
For a calculation of such a characteristic we need to know the full wave functions before and after the emission of photon (which correspond to the unperturbed Hamiltonian). As such a function we shall use general formula (\ref{eq.2.3.4}), where we represent one-nucleon wave functions in the form of the multiplication of space and  spin-isospin functions as
$\psi_{\lambda_{s}} (s) = \varphi_{\lambda_{s}} (\rhobf_{s})\, \bigl|\, \sigma^{(s)} \tau^{(s)} \bigr\rangle$,
%
%
where $\lambda_{s}$ denotes the number of the state of a nucleon with number $s$.
We shall assume that the space function of the nucleon in each state is normalized by the condition
\begin{equation}
  \displaystyle\int |\varphi_{\lambda} (\rhobf_{s})|^{2}\; \mathbf{d} \rhobf_{s} = 1.
\label{eq.app.1.1.3}
\end{equation}
Now we calculate the matrix element (\ref{eq.app.1.1.1}):
\begin{equation}
\begin{array}{lcl}
  Z_{A} (\mathbf{k}) & = &
%
  \displaystyle\frac{1}{A\,(A-1)}
  \displaystyle\sum\limits_{i=1}^{A}
  \displaystyle\sum\limits_{k=1}^{A}
  \displaystyle\sum\limits_{m=1, m \ne k}^{A}
  \biggl\{
    \Bigl\langle \psi_{k}(i)\, \Bigl|\,
      \displaystyle\frac{Z_{i}\,m_{\rm p}}{m_{i}}\; e^{-i \mathbf{k} \cdot \rhobfsm_{i} }\,
    \Bigl|\, \psi_{k}(i)\, \Bigr\rangle\:
    \langle \psi_{m}(j)\, |\, \psi_{m}(j)\, \rangle \quad - \\

  & - &
    \Bigl\langle \psi_{k}(i)\, \Bigl|\,
      \displaystyle\frac{Z_{i}\,m_{\rm p}}{m_{i}}\;  e^{-i \mathbf{k}  \cdot \rhobfsm_{i} }\,
    \Bigl|\, \psi_{m}(i)\, \Bigr\rangle
    \langle \psi_{m}(j)\, |\, \psi_{k}(j)\, \rangle
  \biggr\}.
\end{array}
\label{eq.app.1.1.4}
\end{equation}
Taking into account the orthogonality between wave functions
$\langle \psi_{k}(j)\, |\, \psi_{m}(j)\, \rangle = \delta_{mk}$,
%
%
we obtain
\begin{equation}
\begin{array}{lcl}
  Z_{A} (\mathbf{k}) & = &
%
  \displaystyle\frac{1}{A}
  \displaystyle\sum\limits_{i=1}^{A}
  \displaystyle\sum\limits_{k=1}^{A}\,
    \Bigl\langle \psi_{k}(i)\, \Bigl|\,
      \displaystyle\frac{Z_{k}\,m_{\rm p}}{m_{k}}\;  e^{-i \mathbf{k} \cdot \rhobfsm_{i} }\,
    \Bigl|\, \psi_{k}(i)\, \Bigr\rangle.
\end{array}
\label{eq.app.1.1.6}
\end{equation}
%
%
Taking into account zero charge of neutron, we sum Eq.~(\ref{eq.app.1.1.6}) over spin-isospin states.
For even-even nuclei we obtain
\begin{equation}
\begin{array}{lcl}
  Z_{\rm A} (\mathbf{k}) & = &
  \displaystyle\frac{2}{A}
  \displaystyle\sum\limits_{i=1}^{A}
  \displaystyle\sum\limits_{k=1}^{B}\,
    \langle \varphi_{k}(\rhobf_{i})\, |\,
      e^{-i \mathbf{k} \cdot \rhobfsm_{i}}\,
    |\, \varphi_{k}(\rhobf_{i})\, \rangle,
\end{array}
\label{eq.app.1.1.7}
\end{equation}
where $B$ is the number of states of the space wave function of the nucleon.
Taking into account spin-isospin states, we obtain $B = A / 4$.
%


We define the space wave function of one nucleon in the gaussian form as
\begin{equation}
  \varphi_{i} (\mathbf{r}) =
  N_{x}\,N_{y}\,N_{z} \;
  \exp{\Bigl[-\,\displaystyle\frac{1}{2}\,\Bigl(\displaystyle\frac{x^{2}}{a^{2}} +
    \displaystyle\frac{y^{2}}{b^{2}} + \displaystyle\frac{z^{2}}{c^{2}}\Bigr) \Bigr]} \;
  H_{n_{x}} \Bigl(\displaystyle\frac{x}{a} \Bigr)\,
  H_{n_{y}} \Bigl(\displaystyle\frac{y}{b} \Bigr)\,
  H_{n_{z}} \Bigl(\displaystyle\frac{z}{c} \Bigr),
\label{eq.app.1.2.1}
\end{equation}
where $H_{n_{x}}$, $H_{n_{y}}$ and $H_{n_{z}}$ are the Hermitian polynomials,
$N_{x}$, $N_{y}$, $N_{z}$ are the normalized coefficients.
The unknown normalized coefficients are calculated from the normalization condition:
\begin{equation}
\begin{array}{lcl}
  \vspace{2mm}
  \displaystyle\int
    \Bigl| N_{s}\,
    \exp{\Bigl[-\,\displaystyle\frac{s^{2}}{2a_{s}^{2}} \Bigr]} \;
    H_{n_{s}} \Bigl(\displaystyle\frac{s}{a_{s}} \Bigr)\, \Bigl|^{2}\; ds = 1,
\end{array}
\label{eq.app.1.2.2}
\end{equation}
where $s = x,y,z$.
Taking the properties of the Hermitian polynomials into account (see \cite{Landau.v3.1989}, p.~749),
%
%
we obtain:
\begin{equation}
\begin{array}{ccc}
  N_{x} = \displaystyle\frac{1}{\pi^{1/4} \sqrt{ a\, 2^{n_{x}}\, n_{x}! }}, &
  N_{y} = \displaystyle\frac{1}{\pi^{1/4} \sqrt{ b\, 2^{n_{y}}\, n_{y}! }}, &
  N_{z} = \displaystyle\frac{1}{\pi^{1/4} \sqrt{ c\, 2^{n_{z}}\, n_{z}! }}.
\end{array}
\label{eq.app.1.2.4}
\end{equation}

\subsection{Calculations of the form factor of the nucleus
\label{sec.app.1.3}}

Substituting the one-nucleon space wave function (\ref{eq.app.1.2.1}) into Eq.~(\ref{eq.app.1.1.7}),
we find the form factor for the nucleus:
\begin{equation}
\begin{array}{lcl}
  Z_{\rm A} (\mathbf{k}) & = &
%
%
  \displaystyle\frac{2}{A}\;
  \displaystyle\sum\limits_{i=1}^{A}
  \displaystyle\sum\limits_{n_{x}, n_{y}, n_{z}}^{B}
    I_{x}(n_{x})\, I_{y}(n_{y})\, I_{z}(n_{z}),
\end{array}
\label{eq.app.1.3.1}
\end{equation}
where
\begin{equation}
\begin{array}{lcl}
  I_{x} & = &
  N_{x}^{2} \;
    \exp{\Bigl[-\, a^{2} k_{x}^{2}/4 \Bigr]} \;
  \displaystyle\int\:
    \exp{\Bigl[-\,\displaystyle\frac{(x_{i} + i\,a^{2} k_{x}/2)^{2} }{a^{2}} \Bigr]} \;
    H_{n_{x}}^{2} \Bigl(\displaystyle\frac{x_{i}}{a} \Bigr)\; dx_{i}
\end{array}
\label{eq.app.1.3.2}
\end{equation}
and solutions for $I_{y} (n_{y})$ and $I_{z} (n_{z})$ are obtained after a change of indexes $x \to y$ and $x \to z$.
Let us consider a case of the ground state ($n_{x} = n_{y} = n_{z} = 0$, for example that is for the $\alpha$ particle), where
we have $H_{n_{x}=0} = 1$, $H_{n_{y}=0} = 1$, $H_{n_{z}=0} = 1$.
%
%
In approximation, the integral in Eq.~(\ref{eq.app.1.3.2}) over a complex variable $\tilde{x} = x_{i} + i\,a^{2} k_{x}/2$
has the solution
\begin{equation}
\begin{array}{lcl}
  \displaystyle\int\:
      \exp{\Bigl[-\,\displaystyle\frac{(x_{i} + i\,a^{2} k_{x}/2)^{2} }{a^{2}} \Bigr]}\;
    dx_{i} =
    \displaystyle\int\:
      \exp{\Bigl[-\,\displaystyle\frac{x_{i}^{2}}{a^{2}} \Bigr]}\;
    dx_{i} =
  N_{x}^{-2},
\end{array}
\label{eq.app.1.3.5}
\end{equation}
and we obtain
\begin{equation}
\begin{array}{lcl}
  I_{x} (n_{x}=0)
  & = &
  \exp{\Bigl[-\, a^{2} k_{x}^{2}/4 \Bigr]}.
\end{array}
\label{eq.app.1.3.6}
\end{equation}
In determination of the form factor of the nucleus we have to take into account nonzero states of the one-nucleon space wave function.
We shall find the integral $I_{x}(n_{x} \ne 0)$.
We have
%
%
%
%
%
\begin{equation}
\begin{array}{lcl}
  \displaystyle\int\limits_{-\infty}^{+\infty}
    e^{-(x-y)^{2}}\, H_{n}^{2}(x)\; dx =
  2^{n} \sqrt{\pi}\, n!\, L_{n} (-2y^{2}),
\end{array}
\label{eq.app.1.3.8}
\end{equation}
where $L_{n}$ is the Rodrigues polynomial. 
%
%
For computer calculations the following recurrent formula is useful:
\begin{equation}
\begin{array}{lclll}
  L_{k+1} (x) =
  \displaystyle\frac{1}{k+1}\,
  \Bigl[(2k+1-x)\, L_{k} (x) - k\, L_{k-1} (x) \Bigr] &
  \mbox{\rm at } k \ge 1, &
  L_{0} (x) = 1, &
  L_{1} (x) = 1 - x.
\end{array}
\label{eq.app.1.3.10}
\end{equation}
%
%
%
Using formulas (\ref{eq.app.1.3.8}) and (\ref{eq.app.1.3.10}),
the normalized solution (\ref{eq.app.1.2.4}) for factor $N_{x}$,
we find integral (\ref{eq.app.1.3.2})
%
%
%
\begin{equation}
\begin{array}{lcl}
  I_{x}
  & = &
  L_{n_{x}} \Bigl[a^{2} k_{x}^{2}/2\Bigr] \;
  \exp{\Bigl[-\, a^{2} k_{x}^{2}/4 \Bigr]}
\end{array}
\label{eq.app.1.3.13}
\end{equation}
and calculate the form factor
\begin{equation}
\begin{array}{lcl}
  Z_{\rm A} (\mathbf{k}) & = &
  2\, e^{-\, (a^{2} k_{x}^{2} + b^{2} k_{y}^{2} + c^{2} k_{z}^{2})\,/4}\;
  f_{1}\, (\mathbf{k}, n_{1} \ldots n_{\rm A}),
\end{array}
\label{eq.app.1.3.15}
\end{equation}
where
\begin{equation}
\begin{array}{lcl}
  f_{1}\, (\mathbf{k}, n_{1} \ldots n_{\rm A}) & = &
  \displaystyle\sum\limits_{n_{x}, n_{y},n_{z} = 0}^{n_{x} + n_{y} + n_{z} \le N}
    L_{n_{x}} \Bigl[a^{2} k_{x}^{2}/2\Bigr]\:
    L_{n_{y}} \Bigl[b^{2} k_{y}^{2}/2\Bigr]\:
    L_{n_{z}} \Bigl[c^{2} k_{z}^{2}/2\Bigr].
\end{array}
\label{eq.app.1.3.16}
\end{equation}
Here, function $f_{1}$ is a summation over all states of the one-nucleon space wave function.
Also we use a condition that the form factor tends to the electromagnetic charge of the nucleus at tending
energy of photon to zero:
\begin{equation}
\begin{array}{lcl}
  Z_{\rm A} (\mathbf{k}) \to Z_{\rm A} &
  \mbox{\rm at } k \to 0.
\end{array}
\label{eq.app.1.3.17}
\end{equation}

\section{Emission formed by relative displacements and motions of nucleons inside nucleus
\label{sec.app.2}}


We shall find more accurate approximation of the effective charge (\ref{eq.2.5.5}) than Eq.~(\ref{eq.2.5.6}). Here, we would like to include parameters of the emitted photons into the nuclear form factor.
Rewrite the effective charge (\ref{eq.2.5.5}) as
\begin{equation}
\begin{array}{lcl}
  Z_{\rm eff}^{(2)} (\mathbf{k}, \mathbf{r}) & = &
  \displaystyle\frac{m_{A}\, \tilde{Z}_{\rm p} (\mathbf{k}, \mathbf{r}) -
    m_{\rm p} \tilde{Z}_{A} (\mathbf{k}, \mathbf{r})}{m_{A}+m_{\rm p}},
\end{array}
\label{eq.app.2.1.1}
\end{equation}
where we introduce \emph{extended form factors of the proton and nucleus} as
\begin{equation}
\begin{array}{lcl}
  \tilde{Z}_{\rm p} (\mathbf{k}, \mathbf{r}) =
    z_{\rm p}\; \biggl( e^{i\,\mathbf{k \cdot r}\, \displaystyle\frac{m_{\rm p}}{m_{A}+m_{\rm p}}} - 1 \biggr), &

  \tilde{Z}_{A} (\mathbf{k}, \mathbf{r}) =
    e^{i\,\mathbf{k \cdot r}\, \displaystyle\frac{m_{A} + 2\,m_{\rm p}}{m_{A}+m_{\rm p}}}\,
    Z_{A}(\mathbf{k}) - Z_{A}.
\end{array}
\label{eq.app.2.1.2}
\end{equation}
In further calculation of the matrix element (\ref{eq.2.5.2}) one can join two exponents from the form factors $\tilde{Z}_{\rm p} (\mathbf{k}, \mathbf{r})$ and $\tilde{Z}_{A} (\mathbf{k}, \mathbf{r})$ with an exponent factor $\exp{(-i\, \mathbf{k \cdot r})}$ from the vector potential of the electromagnetic field and then to expand them over multipolar terms. However, such a way requires calculation of a larger number of the radial integrals than, for example, the matrix elements (\ref{eq.2.5.7}) have. Thus, we shall introduce an approximation related to the effective charge.
Let us apply the expansion over the spherical Bessel functions $j_{l} (kr)$,
%
%
%
%
we obtain:
\begin{equation}
\begin{array}{lcl}
  Z_{\rm eff}^{(2)} (\mathbf{k}, \mathbf{r}) =
  \displaystyle\sum\limits_{l=0}^{+\infty}
    i^{l}\, (2l+1)\, P_{l} (\cos \beta)\:
    Z_{\rm eff,\, l}^{(2)} (\mathbf{k}, r)\; - \;
  Z_{\rm eff}^{\rm (dip, 0)},
\end{array}
\label{eq.app.2.1.5}
\end{equation}
where \emph{the partial components of the effective charge} are introduced as
\begin{equation}
\begin{array}{lcl}
  Z_{\rm eff,\, l}^{(2)} (\mathbf{k}, r) & = &
  \displaystyle\frac{m_{A}\, z_{\rm p}}{m_{A}+m_{\rm p}}\;
    j_{l} \Bigl(\displaystyle\frac{m_{\rm p}}{m_{A}+m_{\rm p}}\, kr \Bigr)\; -
  \displaystyle\frac{m_{\rm p}\, Z_{A}(\mathbf{k})}{m_{A}+m_{\rm p}}\,
    j_{l} \Bigl( \displaystyle\frac{m_{A} + 2\,m_{\rm p}}{m_{A}+m_{\rm p}}\, kr \Bigr)
\end{array}
\label{eq.app.2.1.6}
\end{equation}
and $\beta$ is angle between vectors $\mathbf{k}$ and $\mathbf{r}$.
From such a formula one can see that on smaller distances (of variable $r$) the first term should be dominated in the integration of the matrix element, but on far distances the second term (which is decreased more slowly) has a larger contribution.
Such an effective charge should change the shape of the bremsstrahlung spectrum as it changes the dependence of the matrix element on the energy of the photon.



Now we shall consider emission of photons determined by the third matrix element in Eq.~(\ref{eq.2.4.1}).
Performing integration over space variable $\mathbf{R}$, momentum $\mathbf{K}$, we obtain:
\begin{equation}
\begin{array}{lcl}
  \langle \Psi_{f} |\, \hat{H}_{\gamma} |\, \Psi_{i} \rangle_{3} =
  -\,e\; \sqrt{\displaystyle\frac{2\pi\hbar}{w_{\rm ph}}}\,
    \displaystyle\sum\limits_{\alpha=1,2} \mathbf{e}^{(\alpha),*} \cdot
    \Biggl\langle \bar{\Psi}_{f} \Biggl|\,
      e^{i\, \mathbf{k \cdot r}\, \displaystyle\frac{m_{\rm p}}{m_{A}+m_{\rm p}}}\,
      \biggl[
        \displaystyle\sum\limits_{j=1}^{A-1}
          \displaystyle\frac{z_{A j}}{m_{Aj}}\:
          e^{-i \mathbf{k} \cdot \rhobfsm_{Aj}}\,
          \mathbf{\tilde{p}}_{Aj}
      \biggr]\;
    \Biggr|\, \bar{\Psi}_{i} \Biggr\rangle, &
  \mathbf{K}_{i} = \mathbf{K}_{f} + \mathbf{k}.
\end{array}
\label{eq.app.2.2.1}
\end{equation}
Using as functions $\Psi_{\rm p-nucl,\: s} (\mathbf{r})$ the packets (\ref{eq.2.4.13}),
we obtain:
\begin{equation}
\begin{array}{ccl}
  \langle \Psi_{f} |\, \hat{H}_{\gamma} |\, \Psi_{i} \rangle_{3} & = &
  \displaystyle\frac{e}{m}\:
  \sqrt{\displaystyle\frac{2\pi\hbar}{w_{\rm ph}}}\;
  p_{fi,\: 3}\; 2\pi\; \delta(w_{i} - w_{f} - w), \\

  p_{fi,\: 3} & = &
  -\,m\;
    \displaystyle\sum\limits_{\alpha=1,2} \mathbf{e}^{(\alpha),*}\;
  \biggl\langle \psi_{\rm p-nucl, f} (\mathbf{r})\: \biggl|\,
    e^{i\, \mathbf{k \cdot r}\, \displaystyle\frac{m_{\rm p}}{m_{A}+m_{\rm p}}}\,
  \biggr|\, \psi_{\rm p-nucl, i} (\mathbf{r})\: \biggr\rangle \cdot \mathbf{D}_{\rm A} (\mathbf{k}), \\
\end{array}
\label{eq.app.2.2.2}
\end{equation}
where
\begin{equation}
\begin{array}{lcl}
  \mathbf{D}_{\rm A} (\mathbf{k}) & = &
  \Bigl\langle \psi_{\rm nucl, f} (\beta_{A})\: \Bigl|\,
    \displaystyle\sum\limits_{j=1}^{A-1}
      \displaystyle\frac{z_{A j}}{m_{Aj}}\:
      e^{-i \mathbf{k} \cdot \rhobfsm_{Aj}}\,
      \mathbf{\tilde{p}}_{Aj}
  \Bigr|\, \psi_{\rm nucl, i} (\beta_{A})\: \Bigr\rangle.
\end{array}
\label{eq.app.2.2.3}
\end{equation}
Using the expansion
\begin{equation}
\begin{array}{lcl}
  e^{i\, \mathbf{k \cdot r}\, \displaystyle\frac{m_{\rm p}}{m_{A}+m_{\rm p}}} =
  e^{i\, kr\, \cos \beta\; \displaystyle\frac{m_{\rm p}}{m_{A}+m_{\rm p}}} =
  \displaystyle\sum\limits_{l=0}^{+\infty}
    i^{l}\, (2l+1)\, P_{l} (\cos \beta)\:
    j_{l} \Bigl(\displaystyle\frac{m_{\rm p}}{m_{A}+m_{\rm p}} kr \Bigr),
\end{array}
\label{eq.app.2.2.4}
\end{equation}
we rewrite the matrix element as
\begin{equation}
\begin{array}{lcl}
  p_{fi,\: 3} & = &
  -\,\mu\;
  \displaystyle\sum\limits_{l=0}^{+\infty}
    i^{l}\, (2l+1)\, P_{l} (\cos \beta)\; M_{l} (k)\;
  \displaystyle\sum\limits_{\alpha=1,2} \mathbf{e}^{(\alpha),*} \cdot
    \mathbf{D}_{\rm A} (\mathbf{k}),
\end{array}
\label{eq.app.2.2.5}
\end{equation}
where we introduced the nucleon partial matrix elements as
\begin{equation}
\begin{array}{lcl}
  M_{l} (k) & = &
  \biggl\langle \psi_{\rm p-nucl, f} (\mathbf{r})\: \biggl|\,
    j_{l} \Bigl(\displaystyle\frac{m_{\rm p}}{m_{A}+m_{\rm p}} kr \Bigr)\,
  \biggr|\, \psi_{\rm p-nucl, i} (\mathbf{r})\: \biggr\rangle.
\end{array}
\label{eq.app.2.2.6}
\end{equation}
Taking into account the Coulomb gauge and solution (\ref{eq.app.3.10}) for the function $\mathbf{D}_{\rm A} (\mathbf{k})$ given in Appendix \ref{sec.app.3}, we conclude that the matrix element (\ref{eq.app.2.2.5}) and the last matrix element in Eq.~(\ref{eq.2.4.1}) equal zero.


\section{Matrix element over momenta of nucleons of the nucleus
\label{sec.app.3}}

In this appendix we shall calculate the matrix element (\ref{eq.2.7.3}) defined on the basis of
the included operators of momenta of nucleons of the nucleus:
\begin{equation}
\begin{array}{lcl}
  \mathbf{D}_{\rm A} (\mathbf{k}) & = &
  \Bigl\langle \psi_{\rm nucl, f} (\beta_{A})\: \Bigl|\,
    \displaystyle\sum\limits_{j=1}^{A-1}
      \displaystyle\frac{z_{A j}}{m_{Aj}}\:
      e^{-i \mathbf{k} \cdot \rhobfsm_{Aj}}\,
      \mathbf{\tilde{p}}_{Aj}
  \Bigr|\, \psi_{\rm nucl, i} (\beta_{A})\: \Bigr\rangle.
\end{array}
\label{eq.app.3.1}
\end{equation}
Substituting many-nucleon wave function (\ref{eq.2.3.4}),
we calculate it as calculations (\ref{eq.app.1.1.4})--(\ref{eq.app.1.1.7}) and obtain
\begin{equation}
\begin{array}{lcl}
  \mathbf{D}_{\rm A} (\mathbf{k}) & = &
  \displaystyle\frac{1}{A}
  \displaystyle\sum\limits_{i=1}^{A}
  \displaystyle\sum\limits_{k=1}^{A}\,
    \Bigl\langle \psi_{k}(i)\, \Bigl|\,
      \displaystyle\frac{Z_{k}\,m_{\rm p}}{m_{k}}\;  e^{-i \mathbf{k} \cdot \rhobfsm_{i} }\,
      \mathbf{\tilde{p}}_{Aj}
    \Bigl|\, \psi_{k}(i)\, \Bigr\rangle.
\end{array}
\label{eq.app.3.2}
\end{equation}
We sum this expression over spin-isospin states.
In particular, for even-even nuclei we have:
\begin{equation}
\begin{array}{lcl}
  \mathbf{D}_{\rm A} (\mathbf{k}) & = &
  \displaystyle\frac{2}{A}
  \displaystyle\sum\limits_{i=1}^{A}
  \displaystyle\sum\limits_{k=1}^{B}\,
    \langle \varphi_{k}(\rhobf_{i})\, |\,
      e^{-i \mathbf{k} \cdot \rhobfsm_{i}}\,
      \mathbf{\tilde{p}}_{Aj}
    |\, \varphi_{k}(\rhobf_{i})\, \rangle.
\end{array}
\label{eq.app.3.3}
\end{equation}
We substitute the one-nucleon space wave function (\ref{eq.app.1.2.1}) into the matrix element and find
\begin{equation}
\begin{array}{lcl}
  \mathbf{D}_{\rm A} (\mathbf{k}) & = &
%
%
  \displaystyle\frac{2}{A}\;
  \displaystyle\sum\limits_{i=1}^{A}
  \displaystyle\sum\limits_{n_{x}, n_{y}, n_{z}}^{B}
    \Bigl(
      \mathbf{e}_{x}\, J_{x}(n_{x}) +
      \mathbf{e}_{y}\, J_{y}(n_{y}) +
      \mathbf{e}_{z}\, J_{z}(n_{z})
    \Bigr),
\end{array}
\label{eq.app.3.4}
\end{equation}
where orthogonal unit vectors $\mathbf{e}_{\rm x}$, $\mathbf{e}_{\rm y}$,
$\mathbf{e}_{\rm z}$ are used and $\mathbf{e}_{\rm x} = \mathbf{e}^{(1)}$, $\mathbf{e}_{\rm y} = \mathbf{e}^{(2)}$.
Here we introduce a separation on coordinating components $J_{x}(n_{x})$, $J_{y}(n_{y})$, $J_{z}(n_{z})$.
Let us consider the first integral:
\begin{equation}
\begin{array}{lcl}
  J_{x}(n_{x}) & = &
  - i\, \hbar\;
  N_{x}^{2}\;
  \displaystyle\int\,
    \exp{\Bigl[-\,\displaystyle\frac{(x_{i})^{2}}{2a^{2}} \Bigr]} \:
  H_{n_{x}} \Bigl(\displaystyle\frac{x_{i}}{a} \Bigr)\;
    e^{-i k_{x} x_{i}}\;
    \Bigl( \mathbf{e}_{x}\, \displaystyle\frac{d}{d x_{Ai}} \Bigr)\;
    \Bigl\{
      \exp{\Bigl[ -\,\displaystyle\frac{(x_{i})^{2}}{2a^{2}} \Bigr]}\;
      H_{n_{x}} \Bigl(\displaystyle\frac{x_{i}}{a} \Bigr)
    \Bigr\}\;
  dx_{i}\; \times \\

& \times &
  N_{y}^{2}\,
  \displaystyle\int
    \exp{\Bigl[-\,\displaystyle\frac{(y_{i})^{2}}{b^{2}} \Bigr]}\;
  H_{n_{y}}^{2} \Bigl(\displaystyle\frac{y_{i}}{b} \Bigr)\;
    e^{-i k_{y} y_{i}}\;
  dy_{i}\;
  N_{z}^{2}\,
  \displaystyle\int
    \exp{\Bigl[-\,\displaystyle\frac{(z_{i})^{2}}{c^{2}} \Bigr]} \;
  H_{n_{z}}^{2} \Bigl(\displaystyle\frac{z_{i}}{c} \Bigr)\;
    e^{-i k_{z} z_{i}}\;
  dz_{i}.
\end{array}
\label{eq.app.3.5}
\end{equation}
Here the last two integrals represent the found functions $I_{y} (n_{y}, b)$ and $I_{z} (n_{z}, c)$.
We integrate over variable $x$:
\begin{equation}
\begin{array}{lcl}
\vspace{1mm}
  J_{x}(n_{x}) & = &
  \mathbf{e}_{x}\;
  i\, \hbar\;
  N_{x}^{2}
  \displaystyle\int
    \displaystyle\frac{d}{d x_{Ai}}\,
    \Bigl\{
      \exp{\Bigl[-\,\displaystyle\frac{(x_{i})^{2}}{2a^{2}} \Bigr]}\,
      H_{n_{x}} \Bigl(\displaystyle\frac{x_{i}}{a} \Bigr)
    \Bigr\}\;
    e^{-i\, k_{\rm x} x_{i}}\;
    \exp{\Bigl[ -\,\displaystyle\frac{(x_{i})^{2}}{2a^{2}} \Bigr]}\:
    H_{n_{x}} \Bigl(\displaystyle\frac{x_{i}}{a} \Bigr)\;
    dx_{i}\;
    I_{y} (n_{y}, b) \; I_{z} (n_{z}, c)\; + \\

& + &
  \mathbf{e}_{x}\;
  (-i\, k_{\rm x})\:
  i\, \hbar\;
  N_{x}^{2}
  \displaystyle\int
    \exp{\Bigl[-\,\displaystyle\frac{(x_{i})^{2}}{a^{2}} \Bigr]}\,
    e^{-i\, k_{\rm x} x_{i}}\,
    H_{n_{x}}^{2} \Bigl(\displaystyle\frac{x_{i}}{a} \Bigr)\;
    dx_{i}\;
    I_{y} (n_{y}, b)\; I_{z} (n_{z}, c).
\end{array}
\label{eq.app.3.6}
\end{equation}
One can see that the integral over $x$ in the first term is connected with the definition for $J_{x}(n_{x})$,
and the integral over $x$ in the second term to $I_{x} (n_{x}, a)$:
\begin{equation}
\begin{array}{lcl}
  J_{x}(n_{x}) & = &
  -\; J_{x}(n_{x})\; + \;
  \mathbf{e}_{x}\: \hbar\, k_{\rm x}\;
  I_{x} (n_{x}, a)\; I_{y} (n_{y}, b)\; I_{z} (n_{z}, c)
\end{array}
\label{eq.app.3.7}
\end{equation}
and we obtain
\begin{equation}
\begin{array}{lcl}
  J_{x}(n_{x}) & = &
  \mathbf{e}_{x}\:
  \displaystyle\frac{\hbar\, k_{\rm x}}{2}\;
  I_{x} (n_{x}, a)\; I_{y}\; (n_{y}, b)\; I_{z} (n_{z}, c).
\end{array}
\label{eq.app.3.8}
\end{equation}
Now we calculate the function $\mathbf{D}_{\rm A} (\mathbf{k})$ and obtain
\begin{equation}
\begin{array}{lcl}
  \mathbf{D}_{\rm A} (\mathbf{k}) & = &
  \displaystyle\frac{\hbar}{2}\;
  \mathbf{k}\;
  Z_{\rm A} (\mathbf{k}).
\end{array}
\label{eq.app.3.10}
\end{equation}
At the tending energy of the photon to zero ($\mbox{\rm at } k \to 0$),
the matrix element $\mathbf{D}_{\rm A}$ tends to zero.
%

\section{Radial and angular integrals for matrix elements
\label{sec.app.4}}

In this appendix we add results of calculations of the radial and angular integrals for the matrix elements
(\ref{eq.2.8.4})--(\ref{eq.2.8.6}) and (\ref{eq.2.9.7}).
For the first matrix elements (\ref{eq.2.8.4})--(\ref{eq.2.8.6}) we obtain
\begin{equation}
\begin{array}{lcl}
\vspace{1mm}
  p_{l_{\rm ph,\mu}}^{M m_{i} m_{f}} & = &
    \sqrt{\displaystyle\frac{l_{i}}{2l_{i}+1}}\:
      I_{M}^{(m_{i}\, m_{f})} (l_{i},l_{f}, l_{\rm ph}, l_{i}-1, \mu)\;
      \Bigl\{
        J_{1}(l_{i},l_{f},l_{\rm ph}) + (l_{i}+1) \; J_{2}(l_{i},l_{f},l_{\rm ph})
      \Bigr\}\; - \\
\vspace{3mm}
  & - &
    \sqrt{\displaystyle\frac{l_{i}+1}{2l_{i}+1}}\:
      I_{M}^{(m_{i}\, m_{f})} (l_{i},l_{f}, l_{\rm ph}, l_{i}+1, \mu) \;
      \Bigl\{
        J_{1}(l_{i},l_{f},l_{\rm ph}) - l_{i} \; J_{2}(l_{i},l_{f},l_{\rm ph})
      \Bigr\}, \\
\end{array}
\label{eq.app.4.1}
\end{equation}
\begin{equation}
\begin{array}{lcl}
\vspace{1mm}
  p_{l_{\rm ph,\mu}}^{E m_{i} m_{f}} & = &
    \sqrt{\displaystyle\frac{l_{i}\,(l_{\rm ph}+1)}{(2l_{i}+1)(2l_{\rm ph}+1)}} \;
      I_{E}^{(m_{i}\, m_{f})} (l_{i},l_{f}, l_{\rm ph}, l_{i}-1, l_{\rm ph}-1, \mu) \;
      \Bigl\{
        J_{1}(l_{i},l_{f},l_{\rm ph}-1)\; +
        (l_{i}+1) \; J_{2}(l_{i},l_{f},l_{\rm ph}-1)
      \Bigr\}\; - \\
\vspace{1mm}
    & - &
    \sqrt{\displaystyle\frac{l_{i}\,l_{\rm ph}}{(2l_{i}+1)(2l_{\rm ph}+1)}} \;
      I_{E}^{(m_{i}\, m_{f})} (l_{i},l_{f}, l_{\rm ph}, l_{i}-1, l_{\rm ph}+1, \mu) \;
      \Bigl\{
        J_{1}(l_{i},l_{f},l_{\rm ph}+1)\; +
        (l_{i}+1) \; J_{2}(l_{i},l_{f},l_{\rm ph}+1)
      \Bigr\}\; + \\
\vspace{1mm}
  & + &
    \sqrt{\displaystyle\frac{(l_{i}+1)(l_{\rm ph}+1)}{(2l_{i}+1)(2l_{\rm ph}+1)}} \;
      I_{E}^{(m_{i}\, m_{f})} (l_{i},l_{f},l_{\rm ph}, l_{i}+1, l_{\rm ph}-1, \mu) \;
      \Bigl\{
        J_{1}(l_{i},l_{f},l_{\rm ph}-1)\; -
        l_{i} \; J_{2}(l_{i},l_{f},l_{\rm ph}-1)
      \Bigr\}\; - \\
  & - &
    \sqrt{\displaystyle\frac{(l_{i}+1)\,l_{\rm ph}}{(2l_{i}+1)(2l_{\rm ph}+1)}} \;
      I_{E}^{(m_{i}\, m_{f})} (l_{i},l_{f}, l_{\rm ph}, l_{i}+1, l_{\rm ph}+1, \mu) \;
      \Bigl\{
        J_{1}(l_{i},l_{f},l_{\rm ph}+1)\; -
        l_{i} \; J_{2}(l_{i},l_{f},l_{\rm ph}+1)
      \Bigr\},
\end{array}
\label{eq.app.4.2}
\end{equation}
\begin{equation}
\begin{array}{lcl}
\vspace{1mm}
  \breve{p}_{l_{\rm ph,\mu}}^{M m_{i} m_{f}} & = &
    \sqrt{\displaystyle\frac{l_{i}}{2l_{i}+1}}\:
      I_{M}^{(m_{i}\, m_{f})} (l_{i},l_{f}, l_{\rm ph}, l_{i}-1, \mu) \;
      \Bigl\{
        J_{3}(l_{i},l_{f},l_{\rm ph}) + (l_{i}+1) \; J_{4}(l_{i},l_{f},l_{\rm ph})
      \Bigr\}\; - \\
\vspace{3mm}
  & - &
    \sqrt{\displaystyle\frac{l_{i}+1}{2l_{i}+1}}\:
      I_{M}^{(m_{i}\, m_{f})} (l_{i},l_{f}, l_{\rm ph}, l_{i}+1, \mu) \;
      \Bigl\{
        J_{3}(l_{i},l_{f},l_{\rm ph}) - l_{i} \; J_{4}(l_{i},l_{f},l_{\rm ph})
      \Bigr\}, \\
\end{array}
\label{eq.app.4.3}
\end{equation}
\begin{equation}
\begin{array}{lcl}
\vspace{1mm}
  \breve{p}_{l_{\rm ph,\mu}}^{E m_{i} m_{f}} & = &
    \sqrt{\displaystyle\frac{l_{i}\,(l_{\rm ph}+1)}{(2l_{i}+1)(2l_{\rm ph}+1)}} \;
      I_{E}^{(m_{i}\, m_{f})} (l_{i},l_{f}, l_{\rm ph}, l_{i}-1, l_{\rm ph}-1, \mu) \;
      \Bigl\{
        J_{3}(l_{i},l_{f},l_{\rm ph}-1)\; +
        (l_{i}+1) \; J_{4}(l_{i},l_{f},l_{\rm ph}-1)
      \Bigr\}\; - \\
\vspace{1mm}
    & - &
    \sqrt{\displaystyle\frac{l_{i}\,l_{\rm ph}}{(2l_{i}+1)(2l_{\rm ph}+1)}} \;
      I_{E}^{(m_{i}\, m_{f})} (l_{i},l_{f}, l_{\rm ph}, l_{i}-1, l_{\rm ph}+1, \mu) \;
      \Bigl\{
        J_{3}(l_{i},l_{f},l_{\rm ph}+1)\; +
        (l_{i}+1) \; J_{4}(l_{i},l_{f},l_{\rm ph}+1)
      \Bigr\}\; + \\
\vspace{1mm}
  & + &
    \sqrt{\displaystyle\frac{(l_{i}+1)(l_{\rm ph}+1)}{(2l_{i}+1)(2l_{\rm ph}+1)}} \;
      I_{E}^{(m_{i}\, m_{f})} (l_{i},l_{f},l_{\rm ph}, l_{i}+1, l_{\rm ph}-1, \mu) \;
      \Bigl\{
        J_{3}(l_{i},l_{f},l_{\rm ph}-1)\; -
        l_{i} \; J_{4}(l_{i},l_{f},l_{\rm ph}-1)
      \Bigr\}\; - \\
  & - &
    \sqrt{\displaystyle\frac{(l_{i}+1)\,l_{\rm ph}}{(2l_{i}+1)(2l_{\rm ph}+1)}} \;
      I_{E}^{(m_{i}\, m_{f})} (l_{i},l_{f}, l_{\rm ph}, l_{i}+1, l_{\rm ph}+1, \mu) \;
      \Bigl\{
        J_{3}(l_{i},l_{f},l_{\rm ph}+1)\; -
        l_{i} \; J_{4}(l_{i},l_{f},l_{\rm ph}+1)
      \Bigr\},
\end{array}
\label{eq.app.4.4}
\end{equation}
and
\begin{equation}
\begin{array}{ccl}
  J_{1}(l_{i},l_{f},n) & = &
  \displaystyle\int\limits^{+\infty}_{0}
    \displaystyle\frac{dR_{i}(r, l_{i})}{dr}\: R^{*}_{f}(l_{f},r)\,
    j_{n}(k_{\rm ph}r)\; r^{2} dr, \\

  J_{2}(l_{i},l_{f},n) & = &
  \displaystyle\int\limits^{+\infty}_{0}
    R_{i}(r, l_{i})\, R^{*}_{f}(l_{f},r)\: j_{n}(k_{\rm ph}r)\; r\, dr, \\

  J_{3}(l_{i},l_{f},n) & = &
  \displaystyle\int\limits^{+\infty}_{0}
    \displaystyle\frac{dR_{i}(r, l_{i})}{dr}\: R^{*}_{f}(l_{f},r)\,
    j_{n}(k_{\rm ph}r)\;
    Z_{\rm eff}^{(2)} (\mathbf{k}, r)\; r^{2} dr, \\

  J_{4}(l_{i},l_{f},n) & = &
  \displaystyle\int\limits^{+\infty}_{0}
    R_{i}(r, l_{i})\, R^{*}_{f}(l_{f},r)\: j_{n}(k_{\rm ph}r)\;
    Z_{\rm eff}^{(2)} (\mathbf{k}, r)\; r\, dr,
\end{array}
\label{eq.app.4.5}
\end{equation}
\begin{equation}
\begin{array}{ccl}
  I_{M}^{(m_{i}\, m_{f})}\, (l_{i}, l_{f}, l_{\rm ph}, l_{1}, \mu) & = &
    \displaystyle\int
      Y_{l_{f}m_{f}}^{*}(\mathbf{n}_{\rm r})\,
      \mathbf{T}_{l_{i}\, l_{1},\, m_{i}}(\mathbf{n}_{\rm r}) \cdot
      \mathbf{T}_{l_{\rm ph}\,l_{\rm ph},\, \mu}^{*}(\mathbf{n}_{\rm r})\; d\Omega, \\

  I_{E}^{(m_{i}\, m_{f})}\, (l_{i}, l_{f}, l_{\rm ph}, l_{1}, l_{2}, \mu) & = &
    \displaystyle\int
      Y_{l_{f}m_{f}}^{*}(\mathbf{n}_{\rm r})\,
      \mathbf{T}_{l_{i} l_{1},\, m_{i}}(\mathbf{n}_{\rm r}) \cdot
      \mathbf{T}_{l_{\rm ph} l_{2},\, \mu}^{*}(\mathbf{n}_{\rm r})\; d\Omega.
\end{array}
\label{eq.app.4.6}
\end{equation}
The radial integrals $J_{1}$, $J_{2}$ and angular integrals are calculated in~\cite{Maydanyuk.2012.PRC} (see Appendix~B in that paper).
For the matrix element (\ref{eq.2.9.7}) we obtain the following internal matrix components
(see Eqs.~(29), (31), (40) and (41) in~Ref.~\cite{Maydanyuk.2012.PRC}):
\begin{equation}
\begin{array}{lcl}
  \tilde{p}_{l_{\rm ph}\mu}^{M} & = &
    \tilde{I}\,(l_{i},l_{f},l_{\rm ph}, l_{\rm ph}, \mu) \; \tilde{J}\, (l_{i},l_{f},l_{\rm ph}), \\
  \tilde{p}_{l_{\rm ph}\mu}^{E} & = &
    \sqrt{\displaystyle\frac{l_{\rm ph}+1}{2l_{\rm ph}+1}}
      \tilde{I}\,(l_{i},l_{f},l_{\rm ph},l_{\rm ph}-1,\mu) \; \tilde{J}\,(l_{i},l_{f},l_{\rm ph}-1) -
    \sqrt{\displaystyle\frac{l_{\rm ph}}{2l_{\rm ph}+1}}
      \tilde{I}\,(l_{i},l_{f},l_{\rm ph},l_{\rm ph}+1,\mu) \; \tilde{J}\,(l_{i},l_{f},l_{\rm ph}+1),
\end{array}
\label{eq.app.4.7}
\end{equation}
and the corresponding radial and angular integrals
(further calculations of the angular integrals are given in Appendix~B in Ref.~\cite{Maydanyuk.2012.PRC})
\begin{equation}
\begin{array}{lcl}
  \tilde{J}\,(l_{i}, l_{f},n) & = &
  \displaystyle\int\limits^{+\infty}_{0}
    R_{i}(r)\, R^{*}_{f}(l,r)\, j_{n} \Bigl(kr \displaystyle\frac{m_{\rm p}}{m_{A}+m_{\rm p}} \Bigr)\; r^{2} dr, \\

  \tilde{I}\,(l_{i}, l_{f}, l_{\rm ph}, n, \mu) & = &
  \xibf_{\mu} \cdot \displaystyle\int
    Y_{l_{i}m_{i}}({\mathbf n}_{\rm r}^{i}) \:
    Y_{l_{f}m_{f}}^{*}({\mathbf n}_{\rm r}^{f}) \:
    \mathbf{T}_{l_{\rm ph} n,\mu}^{*}({\mathbf n}_{\rm ph}) \: d\Omega.
\end{array}
\label{eq.app.4.8}
\end{equation}

\end{document}